\documentclass[a4paper, amsfonts, amssymb, amsmath, reprint, showkeys, nofootinbib, twoside, superscriptaddress, footnotes]{revtex4-1}
\usepackage[english]{babel}
\usepackage[toc,page]{appendix}
\usepackage{float}
\usepackage[utf8]{inputenc}
\usepackage{tikz}
\usepackage[version=4]{mhchem}
\usetikzlibrary{decorations.pathreplacing} 
\usepackage[colorinlistoftodos, color=green!40, prependcaption]{todonotes}
\usepackage{caption}
\usepackage{subcaption}
\usepackage{amsthm}
\usepackage{mathtools}
\usepackage{graphicx}
\usepackage[left=23mm,right=13mm,top=35mm,columnsep=15pt]{geometry} 
\usepackage{adjustbox}
\usepackage{placeins}
\usepackage[T1]{fontenc}
\usepackage{lipsum}
\usepackage{csquotes}
\usepackage{tikz}
\usepackage{float}
\usetikzlibrary{quantikz}
\usepackage{amsthm}

\newtheorem{theorem}{Theorem}[section]
\newtheorem{Corollary}{Corollary}[section]

\newtheorem{definition}{Definition}[section]
\newtheorem{example}{Example}[section]
\usepackage[unicode=true, colorlinks=true]{hyperref} 
\begin{document}
\title{Noncontextual Pauli Hamiltonians}

\author{Alexis Ralli}
\affiliation{Department of Physics and Astronomy, Tufts University, Medford, MA 02155, USA}
\affiliation{Centre for Computational Science, Department of Chemistry, University College London, WC1H 0AJ, United Kingdom}
\author{Tim Weaving}
\affiliation{Centre for Computational Science, Department of Chemistry, University College London, WC1H 0AJ, United Kingdom}
\author{Peter J. Love}
\affiliation{Department of Physics and Astronomy, Tufts University, Medford, MA 02155, USA}
\affiliation{Computational Science Initiative, Brookhaven National Laboratory, Upton, NY 11973, USA}

\date{\today}

\begin{abstract}
Contextuality is a key feature of quantum mechanics, and identification of noncontextual subtheories of quantum mechanics is of both fundamental and practical importance. Recently, noncontextual Pauli Hamiltonians have been defined in the setting of variational quantum algorithms. In this work we rigorously establish a number of properties of noncontextual Pauli Hamiltonians. 
We prove that these Hamiltonians can be composed of more Pauli operators than diagonal Hamiltonians. This establishes that noncontextual Hamiltonians are able to describe a greater number of physical interactions. We then show that the eigenspaces admit an efficient classical description.
We analyse the eigenspace of these Hamiltonians and prove that for every eigenvalue there exists an associated eigenvector whose stabilizer rank scales linearly with the number of qubits. 
We prove that further structure in these Hamiltonians allow us to derive where degeneracies in the eigenspectrum can arise. 
We thus open the field to a new class of efficiently simulatable states. 
\end{abstract}

\maketitle

\section{Introduction} \label{sec:introduction}

Quantum computing is currently in the noisy intermediate scale quantum (NISQ) era \cite{preskill2018quantum}. NISQ devices have as many as hundreds of qubits but lack error correction, limiting the depth of quantum circuits that can be executed. Utilization of these NISQ devices has led to an explosion of heuristic algorithms that work towards achieving quantum advantage on a NISQ device. So far, quantum advantage for a useful problem has not been achieved. Most NISQ algorithms are heuristic and are focused on achieving experimental realization. As a result, it is unclear if any algorithmic developments are of broader interest outside the world of NISQ devices. One obstacle to this is algorithmic techniques are often developed concurrently with implementations and hence are reported alongside experimental details and error mitigation strategies, making the literature difficult to parse for those only interested in the algorithmic developments.

In the present paper, we provide a self-contained mathematically rigorous presentation of the definition and properties of non-contextual Pauli Hamiltonians. These ideas were developed in the setting of the contextual-subspace variational quantum eigensolver (CS-VQE) method~\cite{kirby2021contextual}. The main ideas presented here were originally given without proof in a sequence of experimental and theoretical papers defining~\cite{kirby2019contextuality, kirby2020classical}, developing~\cite{weaving2023stabilizer, ralli2023unitary} and applying CS-VQE~\cite{kirby2021contextual, weaving2023benchmarking, weaving2025contextual}. The purpose of this paper is to present these ideas in a setting independent of the CS-VQE method alongside rigorous proofs of the main properties of noncontextual Pauli Hamiltonians. We also connect noncontextual Pauli Hamiltonians to related ideas such as stabilizer rank. The proof we provide are straightforward, but either lacking in the existing literature or appear in apparently unrelated parts of the literature. Our aim is to provide a self-contained introduction to noncontextual Pauli Hamiltonians that may be of interest to those working in areas such as Hamiltonian simulation, Hamiltonian complexity, quantum error correction, adiabatic quantum computation, and beyond. 

Quantum mechanics radically departs from the definite states assigned to systems by classical physics. The maximal information available about a quantum system is entirely encoded in the wavefunction. The apparent necessity to track all amplitudes leads to exponential scaling of direct approaches to simulation of quantum systems by classical computers. It is widely believed, (but not proved) that it is not possible to efficiently simulate arbitrary quantum  systems by any means using classical computers. This implies quantum computers are more powerful than classical computers~\cite{preskill2012quantum}. However, not all quantum systems are classically hard to solve, and quantum computers do not provide an advantage for all problems. The question of which quantum systems can be efficiently simulated classically, and which problems require a quantum computer is therefore closely connected to foundational questions about what distinguishes quantum from classical physics. 

Contextuality is an indicator of nonclassicality and is a resource for quantum computation~\cite{howard2014contextuality, shahandeh2021quantum, bermejo2017contextuality, raussendorf2013contextuality, delfosse2015wigner, mansfield2018quantum, karanjai2018contextuality}, and there have been a variety of approaches to quantify contextuality \cite{grudka2014quantifying, budroni2022kochen, gupta2023quantum, pavicic2023quantum}. Noncontextuality is equivalent to positivity of quasi-probability descriptions of quantum mechanics such as the Wigner function~\cite{spekkens2008negativity,ferrie2008frame,ferrie2009framed}, and can be interpreted in terms of semiclassical formulations of quantum mechanics as the order $\hbar^0$ level of the theory~\cite{kocia2017semiclassical,kocia2017discrete,kocia2017discrete2,kocia2018measurement,kocia2019non,kocia2021stationary}.

Any restricted version of quantum mechanics is called a subtheory. One canonical example of a subtheory of quantum mechanics amenable to efficient classical description is the stabilizer formalism for qubits and qudits~\cite{gottesman1996class,gottesman1997stabilizer,gottesman1998heisenberg,aaronson2004improved}. For the stabilizer formalism on qudits of dimension greater than two, and for Gaussian states in continuous variable systems, the classical subtheory can be constructed by imposition of the uncertainty principle by {\em fiat} in a phase space picture: a process called quasi-quantization~\cite{spekkens2016quasi}. The positivity of the corresponding phase-space picture establishes that these subtheories are {\em noncontextual}~\cite{spekkens2008negativity,ferrie2008frame,ferrie2009framed}.

Stabilizer states form an over-complete basis for all quantum states. This means that an arbitrary quantum state can be expressed as a superposition of stabilizer states in many ways. The minimal number of stabilizer states amongst all such superpositions is the stabilizer rank of the state~\cite{garcia2012efficient, garcia2014geometry,bravyi2016trading,bravyi2019simulation}. Determination of stabilizer rank is in general a difficult problem, however, states of bounded stabilizer rank are of interest because they allow efficient classical simulation using, for example, the Bravyi-Gossett algorithm~\cite{PhysRevLett.116.250501}.

One may also obtain a subtheory by restricting the form of the Hamiltonian. Stabilizer Hamiltonians are one such example, and have eigenstates that are all stabilizer states~\cite{kitaev2003fault,dennis2002topological}. Other restricted Hamiltonian forms are, {\em inter alia}, frustration-free Hamiltonians~\cite{bravyi2015gapped}, commuting Hamiltonians, commuting Pauli Hamiltonians, and stoquastic Hamiltonians~\cite{bravyi2006complexity}. For all such restrictions one may ask whether the eigenstates have an efficient classical representation or other signatures of classicality. For example, stoquastic Hamiltonians have off-diagonal matrix elements that are all negative in some basis, implying that the ground state amplitudes in that basis are all positive and hence stoquastic Hamiltonians do not suffer from the sign problem~\cite{bravyi2006complexity}. Hamiltonian complexity formulates the question of classicality vs quantumness through the LOCAL HAMILTONIAN problem~\cite{kitaev2002classical,kempe2006complexity,cubitt2016complexity,osborne2012hamiltonian}. Which classes of Hamiltonians cause the LOCAL HAMILTONIAN problem to lie in a classical or quantum complexity class has been extensively studied~\cite{bravyi2003commutative,bravyi2006complexity,bravyi2006efficient,bravyi2010complexity,aharonov2011complexity}. New types of Hamiltonians that are classical in some respect, but which do not fall into any of the existing classes discussed above are therefore of broad interest in quantum complexity theory, quantum foundations, quantum simulation and beyond.

In this work, we consider noncontextual Pauli Hamiltonians defined in \cite{kirby2020classical}. Noncontextual Pauli Hamiltonians admit a quasi-quantized model~\cite{spekkens2016quasi, kirby2020classical}. The construction and solution of such quasi-quantized models in the context of the noncontextual variational eigensolver method has been described in~\cite{kirby2020classical, kirby2021contextual, ralli2023unitary, weaving2023stabilizer}. In the present paper we aim to provide a self-contained and rigorous presentation of noncontextual Pauli Hamiltonians and some of their properties that we believe will be of broad interest in quantum simulation and Hamiltonian complexity. The goal is to provide a mathematical framework to describe such Hamiltonians rigorously, independent of the original formulation of these Hamiltonians in the setting of the VQE method~\cite{kirby2019contextuality}. The theorems and proofs here will not be challenging for anyone familiar with the stabilizer formalism. However, together they provide a rigorous framework for understanding noncontextual Pauli Hamiltonians that we hope will lead to wider study.

Previous work by Kirby and Love, showed how to describe epistemic states\footnote{joint probability distributions over the ontic states} (all valid states) for noncontexutal Pauli Hamiltonians.  In this work, we further their result by proving that it is always possible to efficiently define the eigenstates of a noncontextual Pauli Hamiltonian directly rather than having to search over epistemic states. This immediately improves the contextual subspace technique \cite{kirby2021contextual}. We further prove that associated with any particular eigenvalue of a noncontextual Pauli Hamiltonian there is always a stabilizer subspace whose dimension scales linearly with the number of qubits.

We prove that noncontextual Pauli Hamiltonians can be composed of at most $2^{n+1}$ unique Pauli operators vs $2^{n}$ for diagonal Hamiltonians. This shows that a noncontextual Hamiltonian can in principle describe twice as many interactions than a diagonal Hamiltonian, enabling a better description of physical processes within a conceptually classical model. 

In Section \ref{sec:background}, we provide necessary background. In Section \ref{sec:discussion} we summarise and discuss our results. We analyse the graph structure and number of operators in noncontextual Pauli Hamiltonians in Subsections  \ref{sec:noncon_g_struc} and \ref{sec:trade_off_sym} . Subsection \ref{sec:block_diagonal} then shows how noncontextual Pauli Hamiltonians have a specific block diagonal form. Following this, Subsection \ref{sec:nc_state_space} analyses the eigenspace of noncontextual Pauli Hamiltonians and shows how the eigenvalue and eigenspace pairs are obtained. Subsection \ref{sec:nc_stab_rank} proves that for each eigenvalue it is always possible to generate an eigenvector with linear stabilizer rank in the number of qubits.  Finally, in Subsection \ref{sec:multiplicity} we analyse the degeneracy that can occur in the eigenspectrum of arbitrary noncontextual Pauli Hamiltonians.

\section{Background} \label{sec:background}

In this Section we summarise background information underpinnning the results of this paper. First in Subsection \ref{sec:Noncontextuality} we explain the concept of Noncontextuality. Subsection \ref{sec:SR_section} provides a brief review of the stabilizer formalism. Subsection \ref{sec:UP} describes how to rotate a linear combination of anticommuting Pauli operators onto a single Pauli operator. Following this, Subsection \ref{sec:Jprod_section} defines the Jordan product. In Subsection~\ref{sec:SymDef} we discuss symmetries of noncontextual Pauli Hamiltonians. Finally, Subsection \ref{sec:Non_P_H} defines a noncontextual Pauli Hamiltonian.

\subsection{Noncontextuality \label{sec:Noncontextuality}}

In a classical probabilistic theory one may always interpret probabilities as arising from ignorance of a definite underlying state. Such underlying states are the ontology of the theory and are called ontic states. The typical example of an ontic state is the position and momentum of every classical particle in a gas. The Maxwell-Boltzmann velocity distribution arises for a single atom from ignorance of all other particles ontic states. A longstanding question in the foundations of quantum mechanics is whether a probabilistic model, or hidden variable theory, built on some kind of ontic states can reproduce the predictions of quantum mechanics. Such a model would reestablish the notion of a definite underlying reality to fundamental physics.

A hidden variable model for quantum mechanics can be called noncontextual if every ontic state determines the probability of the outcomes of every measurement independently of which other measurements are simultaneously performed~\cite{hofer2022two}. In other words, observables can have a definite value independent of what other observables are measured simultaneously. However, the Bell-Kochen-Specker theorem \cite{bell1966problem, kochen1967problem} proves that hidden-variable theories which can explain the predictions of quantum mechanics in a non-contextual way cannot exist in a Hilbert space of dimension greater than two~\cite{cabello1996bell}.
In other words, any hidden variable model that provides definite results for quantum measurements while reproducing the statistical properties of quantum theory must be contextual. For instance, if operators $A$, $B$, and $C$ satisfy $[A, B] = [A, C] = 0$ and $[B, C] \neq 0$, the result of measuring $A$ depends on whether it is measured alone, with $B$, or with $C$ \cite{peres1991two, Peres2002}. A common example of this phenomenon is the Peres-Mermin square \cite{mermin1990simple, peres1991two,mermin1993hidden}, which is an example of ``strong contextuality" \cite{abramsky2011sheaf, kirby2019contextuality}.

One may still ask the question of whether a noncontextual hidden variable theory exists for a particular subtheory of quantum mechanics. In this work, we consider the case of noncontextual Pauli Hamiltonians where it is possible to assign definite eigenvalues to the set of Pauli operators appearing in the Hamiltonian in a context-independent way without incurring any logical contradictions. This allows a noncontextual model to be written for such Hamiltonians \cite{kirby2020classical}. In the following subsections, we present a mathematical framework used to derive the properties of these Hamiltonians.

\subsection{Stabilizer Formalism \label{sec:SR_section}}

A natural basis for all observables on $n$ qubits are the Pauli operators:
\begin{definition}\label{PauliOperators}
The Pauli operators are $n$-fold tensor products of single qubit Pauli operators $\{I,X,Y,Z\}$. There are $4^n$ distinct Pauli operators.
\end{definition}
The Pauli group is defined as follows:
\begin{definition}\label{PauliGroup}
The Pauli group $\mathcal{P}_{n}$ is the set of all Pauli operators with coefficients $\{-i,+i,-1,+1 \}$. The size of $\mathcal{P}_{n}$ is therefore $|\mathcal{P}_{n}| = 4^{n+1}$ \cite{gottesman1998heisenberg}.
\end{definition}
The Clifford group is the set of operators  that leave the Pauli group fixed under conjugation:
\begin{definition}\label{CliffordGroup}
The Clifford group is the normalizer of the Pauli group \cite{gottesman1998heisenberg}. 
\end{definition}
Definition \ref{CliffordGroup} means conjugation of any element of the Pauli group with an element of the Clifford group returns an element of the Pauli group. The Clifford group is the set of all unitaries that can be generated from any combination of $CNOT$, Hadamard and phase $S$ \cite{gottesman1998heisenberg}. Note all Pauli matrices can be constructed from the phase $S$ and Hadamard $H$ gates, $S^{2} = Z, HS^{2}H=X$, $S H S^{2} H S^{\dagger} = Y$, and thus each Pauli operator is also an element of the Clifford group.

The Pauli group contains many subgroups. Amongst them we are particularly interested in stabilizer groups:
\begin{definition}\label{StabilizerGroup}
A Stabilizer group is a subgroup of the Pauli group ($\mathcal{K} \subset \mathcal{P}_{n}$), where all elements commute and which does not contain the element $- I$ (i.e. $-I \notin \mathcal{K}$) \cite{gottesman1996class, gottesman1998heisenberg}.
\end{definition}
Next, we define the following:
\begin{definition}\label{generators}
A generating set of a group $\mathcal{B}$, is a subset $\mathcal{S} \subseteq \mathcal{B}$ such that every element in $\mathcal{B}$ can be expressed as a combination (under the group operation) of finitely many elements of $\mathcal{S}$ and their inverses. 
\end{definition}
We denote a generating set for a group as `the generators' of the group. The generating set $\mathcal{S}$ of a finite group $\mathcal{B}$ has size at most $\log_{2}(|\mathcal{B}|)$ \cite{aaronson2004improved}.
\begin{example}
$\{ {II, ZI, IZ, ZZ} \}$ is a stabilizer group on two qubits. This group is generated by $\{ ZI, IZ \}$. 
\end{example}
The generators of a Stabilizer group provide an efficient description of the group. Furthermore, since $\mathcal{S}$ is abelian and does not contain $-I$ \cite{gottesman1996class}, any element of $\mathcal{K}$ can be described by a particular combination of generators in $\mathcal{S}$ without needing to specify order.

Related to the stabilizer group is the idea of a stabilizer subspace, which is defined as follows:
\begin{definition}\label{StabilizerSubspace}
A Stabilizer group $\mathcal{K}$ specifies a stabilizer subspace on $n$-qubits, which is defined by all $n$-qubit states $\ket{\psi}$ that satisfy $K_{i} \ket{\psi} = \ket{\psi}$ $\forall K_{i} \in \mathcal{K}$.
\end{definition}
In fact, as all elements of the stabilizer group $\mathcal{K}$ commute, the stabilizer subspace will be fully defined by states satisfying $W_{i} \ket{\psi} = \ket{\psi}$ $\forall W_{i} \in \mathcal{W}$, where $\mathcal{W}$ is the set of generators for $\mathcal{K}$ that has size at most $n$ - i.e. ($|\mathcal{W}| \leq n$). The dimension of the stabilizer subspace acted on by $\mathcal{K}$ is  defined by the number of generators of the group $|\mathcal{W}|$ and is explicitly a Hilbert space of dimension $2^{n-|\mathcal{W}|}$. If $|\mathcal{W}|=n$ then there is only one state that is an eigenvalue one eigenvector of all elements of $\mathcal{W}$ and this is called a stabilizer state.

Following Lemma A.2 in \cite{kirby2021contextual}, we state the following corollary:
\begin{Corollary} \label{single_Q_map}
For an independent set of $M \leq n$ Pauli operators that generate a Stabilizer group, it is possible to efficiently determine the Clifford operator that maps each element of the set under conjugation onto a single-qubit $Z$ operator that each act on a distinct qubit.
\end{Corollary}
This is the fundamental property that underpins the qubit tapering method  \cite{bravyi2017tapering}.

Next, we define stabilizer states in terms of the Clifford operations:
\begin{definition}\label{StabilizerState}
Stabilizer states are those generated by a $n$-qubit Clifford unitary $U$ i.e. $\ket{\phi}= U\ket{0}^{\otimes n}$, where any particular $U$ can be constructed solely from the $CNOT$, Hadamard and phase $S$ gates \cite{aaronson2004improved}.
\end{definition}
A given stabilizer state $\ket{\phi}$ is exactly stabilized by $2^{n}$ Pauli operators (see Theorem 1 \cite{aaronson2004improved}). To specify this stabilizer group and the stabilizer state $\ket{\phi}$ , only $n$ group generators need to be specified, hence providing an efficient description of the state.

Stabilizer states form an over-complete basis in which an arbitrary quantum state can be expanded in many ways. This leads to the notion of the stabilizer rank of an arbitrary quantum state: 
\begin{definition}\label{StabilizerRank}
Given an arbitrary $n$-qubit quantum state $\ket{\psi}$, the stabilizer rank of $\ket{\psi}$ is the smallest integer $\chi$ such that $\ket{\psi}$ can be written as:
\begin{equation}
\begin{aligned}
    \label{eq:stabilizer_rank}
     \ket{\psi} = \sum_{i=1}^{\chi} c_{i} \ket{\phi_{i}}.
     \end{aligned}
\end{equation}
\end{definition}

Here $\ket{\phi_{i}}$ are stabilizer states and $c_{i} \in \mathbb{C}$ are complex coefficients. We note that by definition the stabilizer rank must lie between $1 \leq \chi \leq 2^{n}$ \cite{gottesman1998heisenberg, bravyi2016trading}.

\subsection{Linear Combinations of Anticommuting Pauli Operators \label{sec:UP}}

An important unitary rotation $R$, originally constructed for the unitary partitioning technique \cite{izmaylov2019unitary, zhao2020measurement}, maps a normalized linear combination of pairwise anticommuting Pauli operators onto a single Pauli operator. We denote such an operator as
\begin{equation}
\label{eq:AntiSum}
     \hat O = \sum_{\substack{k=1 \\ P_{k} \in \mathcal{A}}}^{|\mathcal{A}|\leq 2n+1} \beta_{k} P_{k},
\end{equation}
where $\mathcal{A} \subset \mathcal{P}_n$ is a set of Pauli operators satisfying
\begin{subequations}
\begin{equation}
  \{P_{a}, P_{b}\}=0 \: \: \forall a \neq b \text{ and } P_{a}, P_{b} \in \mathcal{A},
\end{equation}
with coefficients
\begin{equation}
  \sum_{k}^{|\mathcal{A}|} |\beta_{k}|^{2}  = 1, \beta_{k} \in \mathbb{R}
\end{equation}
\end{subequations}
Note the maximum size of a set of pairwise anticommuting Pauli operators is $2n+1$ i.e. $|\mathcal{A}| \leq 2n+1$ \cite{raussendorf2020phase}, whereas the maximum size of a set of commuting Pauli operators is $2^{n}$.

We state the following theorem:
\begin{theorem}\label{UP_map}
An $l2$-normalized (Euclidean norm) real linear combination of pairwise anticommuting Pauli operators is unitarily equivalent to a single Pauli operator.
\end{theorem}
\noindent This can be shown through a constructive proof, that is provided by results in \cite{izmaylov2019unitary, zhao2020measurement}. Zhao \textit{et al.} provided two efficient constructions of the unitary that maps such an anticommuting set of Pauli operators onto a single Pauli operator \cite{zhao2020measurement}. We define each as follows:

\begin{definition}\label{seq_rot_definition}
 We define $R_{S}$ to be a sequence of rotations, which is a unitary that maps an anticommuting set of Pauli operators to a single Pauli operator by conjugation. It is defined as \cite{zhao2020measurement}:

 \begin{equation}
\begin{aligned}
    \label{eq:R_S}
     R_{\text{S}} = \prod_{\substack{k \\  \forall P_{k}, P_{w} \in \mathcal{A} \\ \forall k \neq w}}^{|\mathcal{A}|} e^{-i \theta_{nk} (iP_{w}P_{k})} = \sum_{j}^{2^{|\mathcal{A}|-1}} c_{j}P_{j}.
     \end{aligned}
\end{equation}
\end{definition}

\begin{definition}\label{LCU_definition}
 We define $R_{\text{LCU}}$ to be the linear combination of Pauli operators that maps an anticommuting set of Pauli operators to a single Pauli operator by conjugation. It is defined as \cite{zhao2020measurement}:

\begin{equation}
\begin{aligned}
    \label{eq:R_LCU}
     R_{\text{LCU}} = \exp(-i \theta \bigg( \sum_{\substack{k \\  \forall P_{k}, P_{w} \in \mathcal{A} \\ \forall k \neq w}}^{|\mathcal{A}|} i\beta_{k}P_{k}P_{w} \bigg))= \sum_{l}^{|\mathcal{A}|} d_{l}P_{l}.
     \end{aligned}
\end{equation}
\end{definition}

In both cases $P$ are Pauli operators, $\theta_{nk}, \theta \in \mathbb{R}$ are rotation angles and $c_{i}, d_{l} \in \mathbb{C}$ and $\beta_{k} \in \mathbb{R}$ are coefficients. Further details concerning the construction of these operators may be found in \cite{zhao2020measurement, ralli2021implementation, ralli2023unitary}.

In summary, conjugating a real linear combination of anticommuting Pauli operators with $R_{\text{LCU}}$ or $R_{\text{S}}$ results in:

\begin{equation}
\begin{aligned}
    \label{eq:UP_summarry}
     R_{\text{LCU}}^{\dagger} \hat O R_{\text{LCU}} = R_{\text{S}}^{\dagger} \hat O R_{\text{S}} = P.
     \end{aligned}
\end{equation}
In words, mapping a linear combination of anticommuting operators onto a single Pauli operator.

We remark that while the $R_{\text{S}}$ and $R_{\text{LCU}}$ constructions have the same action on $\hat O$, they define different unitaries in general. The distinction arises in the number of terms each might generate under conjugation with an element of the Pauli group; $R_{\text{S}}$ may generate an exponential number of terms, while the number of terms in $R_{\text{LCU}}$ scales quadratically as shown in  \cite{ralli2023unitary}.

\subsection{The Jordan product \label{sec:Jprod_section}}

To construct noncontextual Pauli Hamiltonians the Jordan product is required.
\begin{definition}\label{JordanProd}
 The Jordan product of operators $A$ and $B$ is defined as \cite{mccrimmon2004taste}:
 \begin{equation}
\begin{aligned}
    \label{eq:J_prod}
      A \circ B = \frac{\{ A, B \}}{2}.
     \end{aligned}
\end{equation}
\end{definition}
\noindent where $\{\cdot , \cdot\}$ denotes the anticommutator. For Pauli operators, the Jordan product is equal to the regular matrix product if the operators commute and equal to zero if the operators anticommute. This leads to the following Corollary:
\begin{Corollary}\label{JordanProdCorr}
Under the Jordan-Product, no product of two or more anticommuting Pauli operators may be taken when generating a particular Pauli operator.
\end{Corollary}
Finally, we define:
\begin{definition}\label{IndDef}
A Jordan independent set of Pauli operators contains no operator that can be written as a Jordan product over any subset of operators in the set.
\end{definition}
Note that this implies immediately that generating sets of stabilizer groups are Jordan independent.
\begin{example}
Under the Jordan product $\{X, Y, Z \}$ is Jordan independent, as we cannot use $X$ and $Z$ to generate $Y$ (i.e. $Z\cdot X=iY$) because $\{X,Z\}=0$.
\end{example}

Corollary \ref{JordanProdCorr} is important as it ensures no product of anticommuting generators may be taken when constructing a noncontextual Pauli Hamiltonian that would otherwise lead to measurement inconsistencies. Equivalently, when generating a Pauli operator under the Jordan product only a product of commuting operators can be used. We will show later that no inconsistencies arise due to the fact that commuting operators share common eigenstates unlike anticommuting operators. We remark here that this relates to ideas of inference and when it can be used without resulting in logical inconsistencies.  

This relates to closure under inference:
\begin{definition}\label{IndDef}
\cite[Definition 2]{kirby2021contextual} Given an arbitrary set $\mathcal{S}$ of Pauli operators, closure under inference $\overline{\mathcal{S}}$ of $\mathcal{S}$ is the minimal set of Pauli operators, containing $\mathcal{S}$ as a subset, such that every commuting pair $A,B \in \overline{\mathcal{S}}$, $AB$ is also in $\overline{\mathcal{S}}$.
\end{definition}
As discussed in \cite{kirby2020classical}, in principle we can measure $A$, $B$, and their product $AB$ simultaneously, and so can infer the value of $AB$ from any assignment to $A$ and $B$. This leads to closure under inference. We denote a value assignment as consistent if it respects all such inference rules. In our work, noncontextuality is defined by cases where consistent assignments for the set $\overline{\mathcal{S}}$ exist \cite{kirby2019contextuality, kirby2020classical}.

\subsection{Symmetries \label{sec:SymDef}}
Symmetries occur in most physical problems. For example, in quantum chemistry, when solving the electronic Hamiltonian, it is known that the ground state must have $n$-electrons in it. For example, neutral \ce{H2O} must have $10$-electrons. Therefore, when solving such a problem, it would be wise to only look over solutions that contain $n$-electrons. We can mathematically formalize this idea here.

\begin{definition}\label{IndDef}
A symmetry of a Hermitian operator $H$ is any Hermitian operator $S$ where $[S, H] = 0$.
\end{definition}
We note as a symmetry $S$ commutes with $H$, this implies that $S$ and $H$ must share an eigenbasis. If the symmetry operator is degenerate then the  matrix is said to be `block diagonal' with respect to $S$. Each block arises as the problem can be written as projectors onto the different eigenstates of the symmetry operator with a particular eigenvalue. Otherwise, if the symmetry operator is non-degenerate then the two operators share the same eigenstates. Continuing the example at the beginning of this section, the number operator is a symmetry operator and thus in quantum chemistry the second quantized molecular Hamiltonian is block diagonal with respect to different eigenstates of the number operator. 

We note an important subset of these general symmetries are called $\mathbb{Z}_{k}$ symmetries where $S^{k}=I$. These symmetries map system states around closed cycles of length $k \in \mathbb{N}$. The simplest nontrivial case is $k=2$ that has the consequence that it is simultaneously symmetry (commutes) with each of the constituent Hamiltonian terms when decomposed over the Pauli group. We define such a symmetry as follows:

\begin{definition}\label{IndDef}
A $\mathbb{Z}_2$ symmetry of a Hermitian operator $H$ is an operator $A$ that commutes with $H$ ($[A, H] = 0$) and squares to the identity ($A^{2}=I$). 
\end{definition}
We remark that $A$ generates a group of two elements the identity and the operator itself - i.e. $\{ I, A \}$).
The name $\mathbb{Z}_{2}$ stems from the fact that such a set is isomorphic to the cyclic group of order 2. 

We show that from the definition of $\mathbb{Z}_2$ symmetries, we obtain additional properties if we assume the symmetry is a Pauli operator which is a symmetry of a Pauli Hamiltonian. Given  $H = \sum_{i} c_{i} P_{i}$, $[H,P]=0$ for some Pauli operator $P$ and $P^{2}=I$, we know that $P$ generates a stabilizer group. By Corollary \ref{single_Q_map}, we can find a Clifford operator $C$ that implements the following map $C^{\dagger} P C = I_{0}\otimes I_{1} \otimes \hdots \otimes Z _{j} \otimes \hdots \otimes I_{n-2} \otimes I_{n-1}$, where $j$ is a qubit index. We write this single Pauli operator in short as $\Tilde{Z}_{j}$, where the identity terms are omitted. If we transform the Hamiltonian by the same Clifford unitary, we obtain $H' = C^{\dagger} H C$. We note that $[H', \Tilde{Z}_{j}]=0$ as unitary rotations must preserve commutation relations. We can therefore write:
\begin{equation}
\begin{aligned}
    \label{eq:Z2_proof_P1}
      H \mapsto H' &= C^{\dagger} H C = \sum_{i} c_{i} C^{\dagger} P_{i} C = \sum_{i} c_{i} P_{i}' \\
      &= \sum_{i} c_{i} \bigg(\bigotimes_{k=1}^{j-1} \sigma_{k}^{(i)} \bigg) \otimes \sigma_{j}^{(i)} \otimes \bigg(\bigotimes_{k=j+1}^{n} \sigma_{k}^{(i)} \bigg) \\ 
      &= \sum_{i} c_{i} \bigg( Q_{i} \otimes \sigma_{j}^{(i)} \otimes V_{i} \bigg)
     \end{aligned}
\end{equation}
If we then determine $[H', \Tilde{Z}_{j}]$:
\begin{equation}
\begin{aligned}
    \label{eq:Z2_proof_P2}
      [ H', \Tilde{Z}_{j}] = &H' \Tilde{Z}_{j} - \Tilde{Z}_{j} H' \\
      = &\sum_{i} c_{i} \bigg( Q_{i} \otimes \sigma_{j}^{(i)} \cdot Z_{j} \otimes V_{i} \bigg) \\
       &- \sum_{i} c_{i} \bigg( Q_{i} \otimes Z_{j} \cdot \sigma_{j}^{(i)}   \otimes V_{i} \bigg).
     \end{aligned}
\end{equation}
It is clear that $[H', \Tilde{Z}_{j}] =0 \iff [\sigma_{j}^{(i)},Z_{j}] =0$. This means that $\Tilde{Z}_{j}$ commutes with every term in $H'$ and thus $P$ must commute with every term in $H$. From this result, we state the following:

\begin{Corollary}\label{Z2_PauliCorr}
If a single Pauli operator commutes with a Hermitian operator $H = \sum_{P_{i} \in \mathcal{P}_{n}} c_{i} P_{i}$, then it is a 
$\mathbb{Z}_2$ symmetry of that operator and must commute with every individual operator in $H$.
\end{Corollary}

\subsection{Noncontextual Pauli Hamiltonians \label{sec:Non_P_H}}

We first define a $k$-local Hamiltonian~\cite[Definition 14.3]{kitaev2002classical}):
\begin{definition}
A Hamiltonian $H$ acting on $n$ qubits is a $k$-local Hamiltonian if it can be written
\begin{equation}
H=\sum_j H_j
\end{equation}
where each $H_j$ acts nontrivially on at most $k$ qubits for $k\leq n$.
\end{definition}
A special case of $k$-local Hamiltonians are Pauli Hamiltonians where each $H_j$ is a Pauli operator acting on at most $k$ qubits.

To define what a nonconcontextual Pauli Hamiltonian $H_{\text{nc}}$ is, we first state the following theorem:
\begin{theorem}\label{NonconDefWill}
\cite[Theorem 3]{kirby2019contextuality} For a set $\mathcal{S}$ of Pauli operators, let $\mathcal{T} \subseteq \mathcal{S}$  be the set obtained by removing any operator that commutes with all others in $\mathcal{S}$. Then $\mathcal{S}$ is noncontextual if and only if commutation is an equivalence relation on T.
\end{theorem}
\noindent Here the equivalence relation means commutation is transitive on $\mathcal{T}$. In other words, for $P_{i}, P_{j}, P_{k} \in \mathcal{T}$ if $[P_{i}, P_{j}]=0$ and $[P_{j}, P_{k}]=0$ then we can infer that $[P_{i}, P_{k}]=0$. Note that for general Hamiltonians, constructed as a linear combination of Pauli operators, this will not be true. From Theorem \ref{NonconDefWill}, we write the following Definition:

\begin{definition}\label{DonconDef}
A noncontextual Pauli Hamiltonian is any linear combination of Pauli operators generated from a noncontextual set of Pauli operators under the Jordan product.
\end{definition}
The use of the Jordan product (Definition \ref{DonconDef}) guarantees commutation is transitive in a noncontextual set of Pauli operators. We remark that a noncontextual generating set of Pauli operators will always be Jordan independent, but the converse is not true. For example:
\begin{example}
$\{ {XI, YI, ZI, IX, IY, IZ} \}$ is Jordan independent, but not noncontextual. 
\end{example}

We can now write a general noncontextual Pauli Hamiltonian using Definition \ref{DonconDef}:
\begin{equation}
\begin{aligned}
    \label{eq:Hnc}
     H_{nc} = \sum_{P_{i}\in \mathcal{V}} c_{i} P_{i},
     \end{aligned}
\end{equation}
where $\mathcal{V}$ is known as the `support' of $H_{nc}$ i.e. a noncontextual set of Pauli operators appearing in the Hermitian operator, $P_{i}$ are Pauli operators and $c_{i}$ are real coefficients. In subsequent sections we use the notation $|H_{nc}|$ to mean $|\mathcal{V}|$ or the size (number of Pauli operators) of the set .

By Theorem \ref{NonconDefWill}, any noncontextual Pauli Hamiltonian $H_{\text{nc}}$ must be generated by the following Jordan independent generating set of Pauli operators $\mathcal{R}$:
\begin{equation}
\begin{aligned}
    \label{eq:noncon_generators}
     \mathcal{R} &\equiv \mathcal{G} \cup \underbrace{\{ C_{i} | i=0,1,\hdots, M-1 \}}_{\mathcal{A}},
     \end{aligned}
\end{equation}
as proved in \cite{kirby2020classical}. Here $\mathcal{G}$ is an independent set of Pauli operators that each commute with every Pauli operator in $\mathcal{R}$ (and thus with every operator in $\mathcal{V}$) and the set $\mathcal{A}$ is a set of pairwise anticommuting Pauli operators. We remark that every element in $\mathcal{G}$ will be a $\mathbb{Z}_{2}$ symmetry of the noncontextual Pauli Hamiltonian generated by $\mathcal{R}$. We note that different combinations of operators in the set $\mathcal{G}$ must also produce symmetries of $H_{\text{nc}}$. We therefore sometimes denote $\mathcal{G}$ as a set of symmetry generators.

In other words for an arbitrary noncontextual set of Pauli operators $\mathcal{S}$, $\mathcal{R}$ will be an independent generating set for $\overline{\mathcal{S}}$ under the Jordan product. This allows us to state the following Corollary derived from Theorem \ref{NonconDefWill}:

\begin{Corollary}
 A set of Pauli operators is noncontextual if and only if the independent generating sets for its closure under inference
are composed of a set of universally-commuting operators ($\mathcal{G}$), and a set of pairwise anticommuting operators $\mathcal{A}$
\end{Corollary}
\noindent as discussed in Appendix B of \cite{kirby2020classical}.

In the next Section, we show that this definition of noncontextual Pauli Hamiltonians dictates many of their properties.

\section{Structure of Noncontextual Pauli Hamiltonians} \label{sec:discussion}

In this section, we analyse the allowed structure of noncontextual Pauli Hamiltonians. Subsection \ref{sec:noncon_g_struc} analyzes the structure of the compatibility graph  of $H_{\text{nc}}$. Subsection \ref{sec:trade_off_sym} analyzes the maximum number of Pauli operators allowed in the support of $H_{\text{nc}}$. Subsection \ref{sec:block_diagonal} then shows the block diagonal form of noncontextual Pauli Hamiltonians. In subsection \ref{sec:nc_state_space} we investigate the properties of the eigenspaces of $H_{\text{nc}}$.  Following this, in subsection \ref{sec:nc_stab_rank} we take a closer look at the eigenstates of $H_{\text{nc}}$ and in subsection \ref{sec:multiplicity} we look into the multiplicity of different eigenvalues for a general noncontextual Pauli Hamiltonian.

\subsection{Compatibility Graph of Noncontextual Hamiltonians \label{sec:noncon_g_struc}}

The commutation relations of a set of Pauli operators may be represented by their compatibility graph:
\begin{definition}\label{compgraph} The {\em compatibility graph} of a set of Pauli operators $\{P_i\}$ is the graph $\mathcal{Q} = (\mathcal{V}, \mathcal{E})$ where $\mathcal{V}$ is a set of vertices labelled by operators, and $\mathcal{E} = \{\{i, j\} \in \mathcal{E} \iff [P_i, P_j] = 0 \; \forall \; P_i, P_j \in \mathcal{V} \}$ is the set of edges of the graph.
\end{definition}

Various subgraphs of the compatibility graph are of interest for noncontextual Hamiltonians. We use the convention that the subgraph of a graph $G=(V,E)$ induced by a subset of its vertices $V'$ is denoted $G_{V'}$. We define the subgraph corresponding to $\mathbb{Z}_2$ Pauli symmetries of a set of Pauli operators $\{P_i\}$ as follows:
\begin{definition}
The subgraph $Q_{\mathcal{Z}}$ is induced by the set $\mathcal{Z} \coloneqq \{i \in \mathcal{V} :\{i,j\} \in \mathcal{E} \;\forall j \in \mathcal{V} \}$: the set of vertices connected to all other vertices. 
\end{definition}
The set $\mathcal{Z}$ represents Pauli operators generated from only $\mathcal{G} \subset \mathcal{R}$  (equation \ref{eq:noncon_generators}). The graph $\mathcal{Q}$ describes a noncontextual system of Pauli operators if and only if the subgraph induced by $\mathcal{V} \setminus \mathcal{Z}$ can be partitioned into disjoint \textit{cliques}: complete subgraphs $\mathcal{C}_k$ such that $Q_{\mathcal{V} \setminus \mathcal{Z}} = \cup_{k=1}^{M} \mathcal{C}_k$ and $\mathcal{C}_k \cap \mathcal{C}_l = \emptyset \;\forall k \neq l$. In words, after removing Pauli operators that commute universally (i.e. the set $\mathcal{Z}$) the system is noncontextual if and only if the remaining vertices ($\mathcal{T}$) form a set of disjoint complete subgraphs. Expressed in this way, the identification of noncontextual subsystems is an instance of the $NP$-hard disjoint union of cliques problem \cite{jansen1997disjoint}, as discussed in \cite{kirby2021contextual}.

Figure \ref{fig:noncon_Graph} shows the structure of all possible $4$-qubit noncontextual Hamiltonians.  Here $2 \leq |\mathcal{A}|\leq 2n+1 = 9$. The central clusters represent globally commuting operators and the outer clusters represent disjoint commuting cliques. We note that by removing the globally commuting operators we obtain the disjoint commuting cliques.

We use this fact to state the following corollaries.

\begin{Corollary}\label{graph1}
For an arbitrary set of Pauli operators $\mathcal{S}$, let $\mathcal{Z}$ be the set of operators that can be generated by the $\mathbb{Z}_{2}$ symmetries of $\mathcal{S}$. Then $\mathcal{S}$ is noncontextual if and only if the compatibility graph of the resulting set $\mathcal{S} \setminus \mathcal{Z}$ is comprised of disjoint cliques.
\end{Corollary}

\begin{Corollary}\label{graph2}
For an arbitrary set of Pauli operators $\mathcal{S}$, let $\mathcal{Z}$ be the set of operators which can be generated by the $\mathbb{Z}_{2}$ symmetries of $\mathcal{S}$. Then $\mathcal{S}$ is noncontextual if and only if the anti-compatibility graph of $\mathcal{S} \setminus \mathcal{Z}$ is multipartite complete.
\end{Corollary}
Note the anti-compatibility graph is a graph where edges are between anti-commuting operators.

The significance of Corollary \ref{graph1} and Corollary \ref{graph2} is that they define the structure of noncontextual sets in terms of graphs. This structure could potentially be used to improve the problem of extracting noncontextual sets from a larger contextual set of Pauli operators (arbitrary graph), where currently the only published approach is the greedy heuristic introduced in \cite[Section IV]{kirby2019contextuality}. 

A pertinent question that emerges from this graph structure is: What is the maximum size of a noncontextual set of Pauli operators?

\subsection{Size of Support of Noncontextual Pauli Hamiltonians \label{sec:trade_off_sym}}  

The Pauli operators in $\mathcal{G}$ are an independent generating set for $\mathcal{Z}$, where every element in $\mathcal{Z}$ is a $\mathbb{Z}_2$ symmetry of $H_{nc}$. Hence $\mathcal{G}$ is a generating set for all the $\mathbb{Z}_2$ symmetries of the noncontextual Pauli Hamiltonian. Furthermore, $\mathcal{G}$ forms a generating set for a stabilizer group. The maximum size of $\mathcal{G}$ is therefore $n$: $|\mathcal{G}| \leq n$. In the case $|\mathcal{G}| = n$, there is only one state that has all the symmetries in $\mathcal{Z}$, there can be no other terms in the Hamiltonian and therefore the Hamiltonian is a stabilizer Hamiltonian with at most $2^n$ terms. 

If $|\mathcal{G}|<n$, then the stabilizer subspace described by $\mathcal{G}$ will be of dimension $2^{n-|G|}$. We can use the result of Corollary~\ref{single_Q_map} and assume that $G$ is a set of single qubit $Z$ operators each acting on one of the first $|G|$ qubits. For an $n$-qubit problem there will therefore be an additional subspace defined on $n-|G|$ qubits not acted on by elements of $\mathcal{G}$. Then we can define a set of pairwise anti-commuting Pauli operators $\mathcal{A}$, that act in this subspace while commuting with every element of $\mathcal{G}$. This means they can only act nontrivially on the $n-|G|$ qubits. Then the size of $\mathcal{A}$ can be at most  $|\mathcal{A}| \leq 2 (n - |\mathcal{G}|) + 1$, where $[P_{i}, P_{j}]=0$ $\forall P_{i} \in \mathcal{A}$ $\forall P_{j} \in \mathcal{G}$ and $\{P_{k}, P_{l}\}=0$ $\forall  P_{k},P_{l}  \in \mathcal{A}$ $\forall k \neq l$. This stems from the fact that a set of pairwise anticommuting Pauli operators on $m$ qubits can have size at most $2m+1$. 

\begin{example}
$\mathcal{G} \equiv \{ZXZ,YIY \}$ has size two. This means there will be a subspace defined on $n-|\mathcal{G}|=3-2=1$ qubit not acted on by elements of $\mathcal{G}$. It is possible to therefore define a set $\mathcal{A}$ of size $2(n-|\mathcal{G}|)+1 \leq 3$ pairwise anticommuting Pauli operators such that each element of this set commutes with all elements in $\mathcal{G}$. For example: $\mathcal{A} \equiv \{YYI,ZYX,XIX \}$.
\end{example}

If we begin instead with a set $\mathcal{A}$ of $M$ anticommuting Pauli operators acting on $n$ qubits then we can perform a Clifford rotation to obtain a set of pairwise anticommuting Pauli operators that act nontrivially on at most $\lceil \frac{M-1}{2} \rceil$. In this case the size of the symmetry generating set $\mathcal{G}$ will be upper bounded by $|\mathcal{G}| = n - \lceil \frac{M-1}{2} \rceil$. Following these ideas alongside Theorem \ref{NonconDefWill} and equation \ref{eq:noncon_generators} allows us to state the following Corollary:

\begin{Corollary}\label{NonconMult}
The size of the set of Pauli operators in a noncontextual Pauli Hamiltonian is upper bounded as:
\begin{equation}
\label{eq:NonconTerms}
\begin{aligned}
    |H_{\mathrm{noncon}}|     &\leq 2^{|\mathcal{G}|} + |\mathcal{A}|2^{|\mathcal{G}|}    \\
    &\leq 2^{|\mathcal{G}|} (1+ |\mathcal{A}|)    \\
    &\leq 2^{|\mathcal{G}|} \bigg( 2(n-|\mathcal{G}|) + 2 \bigg)\\
    &\leq 2^{|\mathcal{G}|+1} \big(n-|\mathcal{G}| + 1 \big) \\
    &\leq 2^{n+1}.
\end{aligned}
\end{equation}
\end{Corollary}

Inspecting the right hand side of the first line of equation \ref{eq:NonconTerms}, the first term counts all possible operators that can be generated by products of operators in $\mathcal{G}$. The second term counts the maximum number of operators that can be generated by any combination of Pauli operators in $\mathcal{G}$ and a single element of $\mathcal{A}$. Note this is because under the Jordan product if more than one element of $\mathcal{A}$ is used to generate a Paul operator the product will be zero. The third line of equation \ref{eq:NonconTerms} comes from setting $|\mathcal{A}| = 2 (n - |\mathcal{G}|) + 1$, which is the  maximum number of elements $\mathcal{A}$ can have given $\mathcal{G}$ as discussed in the beginning of this subsection. Further special cases are discussed in Appendix~\ref{sec:nonconHSize}.

In Figure \ref{fig:noncon_Graph}, for $2,3,4$ and $5$ cliques the number of vertices (Pauli operators) is greater than $2^{4}=16$, which represents a Pauli Hamiltonian generated by a set of $n$-independent commuting generators. For the $7$ clique case, we observe that the number of nodes exactly equals $16$. All other graphs have fewer than $16$ Pauli operators. We note the number of terms in a noncontextual Hamiltonian is largest when $|\mathcal{G}|=n-1$ and $|\mathcal{A}|=3$. 

We remark that our analysis here is strictly looking at the maximal number of terms in different noncontextual $H_{\text{nc}}$. However, when studying a physical system the Hamiltonian under consideration will have a polynomial number of terms. A more interesting question to address is what physical interactions a noncontextual Pauli Hamiltonian can describe when composed of only a polynomial number of Pauli operators. However, such an analysis will be system dependent. We leave further analysis in different domains, such as quantum chemistry, condensed matter and high energy physics, to future work.

\begin{figure}[t]
    \includegraphics[width=\linewidth]{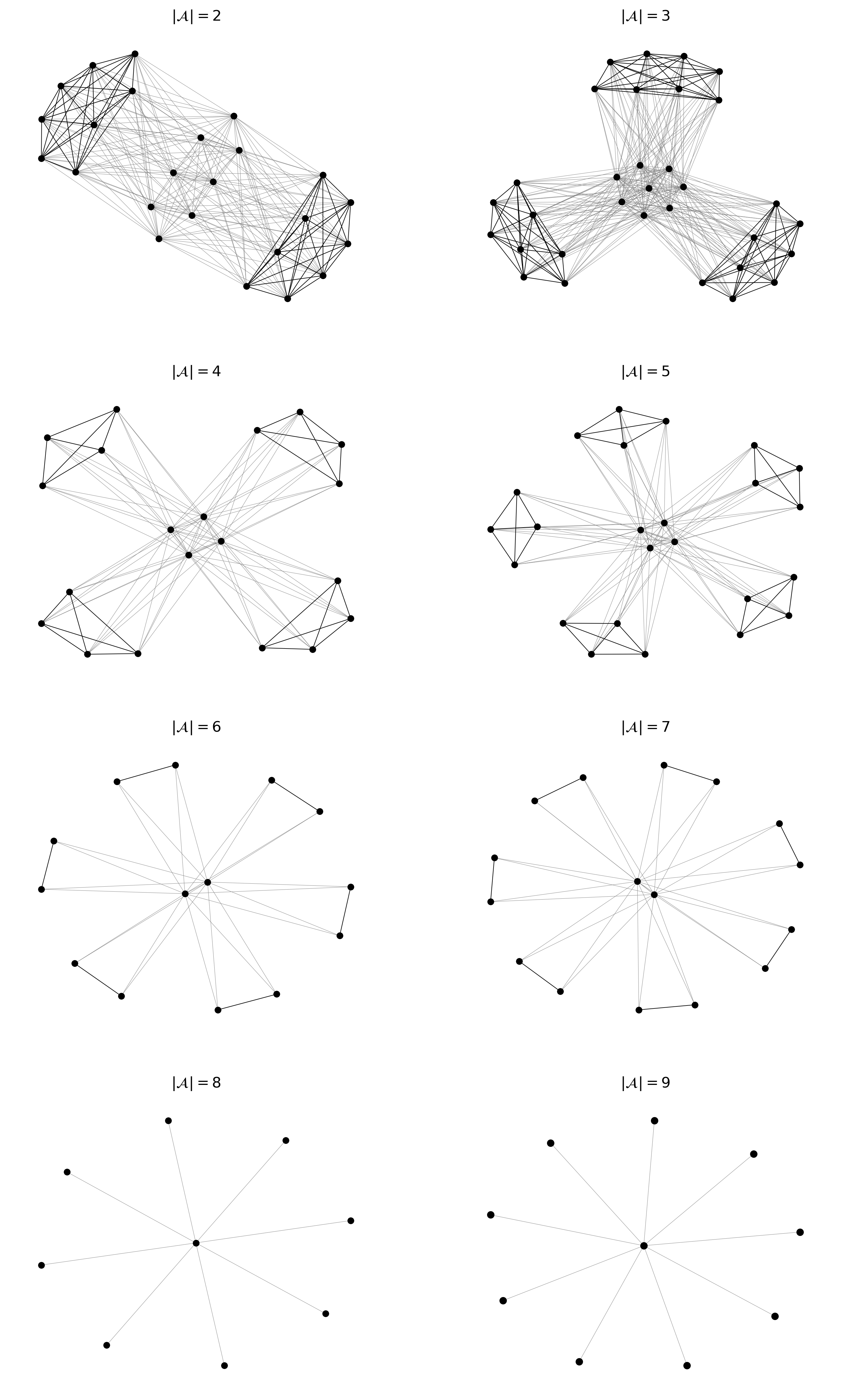}
    \caption{Different $4$-qubit noncontextual Pauli Hamiltonians by size of $\mathcal{A}$. Vertices represent Pauli operators and edges are between commuting Pauli operators.  Each central cluster represents globally commuting Pauli operators ($P \in \mathcal{Z}$) and outer clusters represent disjoint commuting cliques. Note the identity operator is included in these plots.}
    \label{fig:noncon_Graph}
\end{figure}

\subsection{Block Structure of Noncontextual Pauli Hamiltonians \label{sec:block_diagonal}}

Can we ascertain anything about the spectrum and eigenvectors of a general Noncontextual Pauli Hamiltonian $H_{\text{nc}}$? For a general Noncontextual Pauli Hamiltonian every Pauli operator $G_{i} \in \mathcal{G}$ commutes with the Hamiltonian $[G_{i}, H_{nc}] =0$ $\forall G_{i}\in \mathcal{G}$. Each $G_{i}$ therefore represents a  Pauli $\mathbb{Z}_2$ symmetry of $H_{\text{nc}}$, and each eigenstate of $H_{nc}$ must also be an eigenstate of each $G_{i}$. We can therefore block diagonalize $H_{nc}$ where the blocks are labelled by the eigenvalues $\vec \nu$ of the $G_{i}$, where $\nu_{i} = \{+1, -1\}$ gives $\langle G_{i} \rangle$ and $|\vec{\nu}|= |\mathcal{G}|$. The projector onto a block labelled by $\vec \nu$ is given by:
\begin{equation}
\label{eq:projectors}
\begin{aligned}
\Pi(\vec{\nu}) = \frac{1}{2^{|G|}}\prod_{i}^{|G|} \big( I + \nu_{i}G_{i} \big)
\end{aligned}
\end{equation}
Note that the projector in equation \ref{eq:projectors} is a projector onto a stabilizer subspace. 

We can formalize this with the following Theorem:
\begin{theorem}\label{block_theorem}
A noncontextual Pauli Hamiltonian is block diagonal with respect to elements in the set $\mathcal{G}$.
\end{theorem}
\textit{Proof.} We write the noncontextual Hamiltonian in terms of projectors onto the eigenspace of each element of $\mathcal{G}$ (a symmetry):

\begin{equation}
\label{eq:projectors2}
\begin{aligned}
H_{nc} &= \sum_{\vec{\nu} \in \{+1,-1 \}^{\times |\mathcal{G}}} \Pi(\vec{\nu}) \cdot H_{nc} \cdot \Pi(\vec{\nu})^{\dagger} \\
     &=\sum_{\vec{\nu} \in \{+1,-1 \}^{\times |G|}} H_{nc}(\vec{\nu})
\end{aligned}
\end{equation}
with the sum going over all $2^{|G|}$ assignments of $\vec{\nu}$. Here a given $\vec{\nu}$ assignment fixes the expectation value of all operators in $\mathcal{G}$. \qedsymbol{} 

The interpretation of equation \ref{eq:projectors2} is that each projector is acting as a partial protective measurement onto the eigenspace of a symmetry generator in $\mathcal{G}$ \cite{weaving2023stabilizer}. For a specific $\vec{\nu}$ assignment the eigenvalues of all the operators in $\mathcal{Z}$ (i.e. those generated by only operators in $\mathcal{G}$) will be fixed. We note that $\mathcal{G}$ is a Stabilizer group; therefore, each $\vec{\nu}$ defines a stabilizer subspace. For the purposes of determining the spectrum of $H_{nc}$ we can restrict to considering only one block at a time, and therefore finding eigenstates that lie in a particular stabilizer subspace.  We can therefore write out the form of each $H_{nc}(\vec{\nu})$:
\begin{equation}
\begin{aligned}
    \label{eq:vec_assignment}
     H_{\text{nc}} (\vec{\nu}) &= \delta(\vec{\nu}) I + \sum_{\substack{i \\ P_{i} \in \mathcal{A}}} c_{i}(\vec{\nu}) P_{i}
     \end{aligned}
\end{equation}
Here the $\delta(\vec{\nu}) \in \mathbb{R}$ arises from any constant term present in $H_{nc}$ plus a contribution to the energy from all the Pauli operators in $\mathcal{Z}$. This is because all the operators in  $\mathcal{Z}$ can be generated from the operators in $\mathcal{G}$ that expectation value is now fixed by the projector $\Pi(\vec{\nu})$. \qedsymbol{} 

In other words, $H_{nc}$ is block diagonal with respect to $\mathcal{G}$. Furthermore, each $H_{\text{nc}} (\vec{\nu})$ is a linear combination of pairwise anticommuting Pauli operators from the set $\mathcal{A}$ plus a constant shift. 

Equivalently, it is possible to simultaneously diagonalize the operators in $G$ via an efficiently implementable Clifford operation $C$ (see Corollary \ref{single_Q_map}) \cite{bravyi2017tapering,kirby2021contextual, ralli2023unitary, weaving2023stabilizer}. We can write the Hamiltonian under this transformation as:
\begin{equation}
\label{eq:projectors4}
\begin{aligned}
H_{nc} \mapsto& C^{\dagger}H_{nc}C = H_{nc}' \\
&= \begin{pmatrix}
H_{\text{nc}}' (\vec{\nu}_{1}) & \mathbf{0}  & \cdots  & \mathbf{0} \\ 
\mathbf{0} & H_{\text{nc}}' (\vec{\nu}_{2}) & \cdots & \mathbf{0} \\ 
\vdots  & \vdots & \ddots  & \vdots \\ 
\mathbf{0} & \mathbf{0}  & \cdots & H_{\text{nc}}' (\vec{\nu}_{2^{G}})
\end{pmatrix}.
\end{aligned}
\end{equation}
where $H_{\text{nc}}' (\vec{\nu}_{j}) = C^{\dagger}H_{nc}(\vec{\nu}_{j})C$. Without loss of generality, describing $H_{nc}$ in this basis is useful, because conjugating each symmetry generator $G_{i} \in \mathcal{G}$ with $C$ results in each element of $\mathcal{G}$ being mapped onto a single qubit Pauli $Z$ acting on a unique qubit \cite{bravyi2017tapering}. This simplifies the definition of the projectors defined in equation \ref{eq:projectors}. Now projecting onto an eigenstate of a particular $C^{\dagger} G_{i}C$ symmetry is equivalent to fixing the unique qubit it acts on to be in the state $\ket{0}$ or $\ket{1}$. For each symmetry, the $+1$ or $-1$ eigenspace can be selected and after projection the remaining Hamiltonian will be defined on one less qubit. While this is equivalent to equation \ref{eq:projectors2}, in that scenario the symmetry generators are not necessarily single qubit Pauli matrices and thus the state the qubits will be fixed to under a projection onto a given eigenspace is more complicated.

In the next subsection, we show that it is efficient to classically diagonalize any block of $ H_{\text{nc}} (\vec{\nu})$ of $H_{nc}$.

\subsection{Noncontextual Spectrum\label{sec:nc_state_space}}

Given a particular block of a noncontextual Hamiltonian, $H_{\text{nc}} (\vec{\nu})$ (equation \ref{eq:vec_assignment}), we state the following theorem:
\begin{theorem}\label{AntiCommExpec}
Any $n$-qubit Hermitian operator defined by a linear combination of pairwise anticommuting Pauli operators will have two distinct eigenvalues $\pm \|\vec{c}\|_{2}$, where $\vec{c}$ is the vector of coefficients in the linear combination. Here $\|\vec{c}\|_{2}$ is the Euclidean norm of $\vec{c}$. Furthermore, each eigenvalue will occur $2^{n-1}$ times meaning the eigenspace is equally split into a positive and negative subspace.
\end{theorem}
We provide a proof of this in Appendix \ref{sec:AC_exp}. Re-writing equation \ref{eq:vec_assignment}, by normalizing the sum, we obtain:
\begin{equation}
\begin{aligned}
    \label{eq:vec_assignment2}
     H_{\text{nc}} (\vec{\nu}) &= \delta(\vec{\nu}) I +  \|\vec{c}(\vec{\nu})\|_{2} \sum_{\substack{i \\ P_{i} \in \mathcal{A}}} \beta_{i} P_{i}.
     \end{aligned}
\end{equation}
where $\beta_{i} = \frac{c_{i}(\vec{\nu})}{(\sum_{i}c_{i}(\vec{\nu})^{2})^{1/2}}= \frac{c_{i}}{\|\vec{c}(\vec{\nu})\|_{2}}$. By Theorem \ref{AntiCommExpec}, we can write the eigenvalues of $H_{\text{nc}} (\vec{\nu})$ or noncontextual energies as:

\begin{equation}
\begin{aligned}
\label{eq:vec_assignment_energy}
   E_\pm(\vec{\nu}) = \delta(\vec{\nu}) \: I \pm \bigg(\sum_{i} |c_{i}(\vec{\nu})|^{2}\bigg)^{1/2},
\end{aligned}
\end{equation} 
where the negative sign always produces the lowest energy; an alternative derivation of this expression, which is more convenient for implementation is given in Appendix \ref{noncon_obj_func}. A consequence of equation \ref{eq:vec_assignment_energy} is each $H_{\text{nc}} (\vec{\nu})$ block in $H_{nc}$, equation \ref{eq:projectors2}, will be split into a positive and negative subspace.

The significance of Theorem \ref{block_theorem} and Theorem \ref{AntiCommExpec} is that eigenvalues of  arbitrary noncontextual Pauli Hamiltonians are efficient to classically compute given a particular binary assignment $\vec{\nu}$. This formulation improves upon prior work \cite{kirby2020classical,kirby2021contextual,weaving2023stabilizer,ralli2023unitary}, where a convex optimization (over a unit vector) has now been removed. The ground state of $H_{\text{nc}}$ is therefore obtained by finding $\vec{\nu}$ that gives the lowest $E(\vec{\nu})$. 

It is important to note that there are $2^{|\mathcal{G}|}$ unique choices for $\vec{\nu}$ and thus $2^{|\mathcal{G}|}$ blocks in $H_{\text{nc}}'$ and so in general it would take exponential time to solve all the blocks as the size of $\mathcal{G}$ scales as $\mathcal{O}(n)$. Finding the noncontextual ground state is therefore an NP-complete problem, consistent with the proof in \cite{kirby2020classical}. 

In the next section, we will show how to construct the eigenstates corresponding to each eigenvalue $E(\vec{\nu})$.

\subsection{Stabilizer Rank of Noncontextual Eigenstates \label{sec:nc_stab_rank}}

Given a particular noncontextual eigenvalue $E_\pm(\vec{\nu})$, specified by the assignment $\vec{\nu}$, how can we generate the associated eigenspace? Clearly, by construction any eigenvector corresponding to a particular $\vec{\nu}$ must be stabilized by all $\nu_{i} G_{i}$ and $\pm P_{\mathcal{A}}$. Here $P_{\mathcal{A}}$ is a single Pauli operator obtained by conjugating $H_{\text{nc}} (\vec{\nu}) $ (equation \ref{eq:vec_assignment2}) with $R$ (Section \ref{sec:UP}) ignoring the identity operator. As $R$ will be defined by the coefficients in $H_{\text{nc}} (\vec{\nu})$ we denote $R$ for a $H_{\text{nc}} (\vec{\nu}) $ as $R(\vec{\nu})$.
Without loss of generality, the independent set $\{ \nu_{i} G_{i} | \forall G_{i} \in \mathcal{G}\} \cup \{\pm P_{\mathcal{A}}\}$ defines a stabilizer subspace and also generates a stabilizer group. Any state in this stabilizer subspace will be a valid noncontextual eigenstate with eigenvalue $E_\pm(\vec{\nu})$. We can write a general form for a noncontextual eigenstate $\ket{\psi_\pm^{\text{nc}}(\vec{\nu})}$ with eigenvalue $E_\pm(\vec{\nu})$ as:

\begin{equation}
 \label{eq:noncon_gs}
\begin{aligned}
    \ket{\psi_\pm^{\text{nc}}(\vec{\nu})} = R^{\dagger}(\vec{\nu})\ket{\psi_\pm(\vec{\nu})}.
\end{aligned}
\end{equation}
We note that $\ket{\psi_\pm(\vec{\nu})}$ is a stabilizer state and has an efficient classical description as it is fully specified by $\mathcal{G}$ and $\vec{\nu}$ that each have size at most $n$, and the sign of $\pm P_\mathcal{A}$.

In Subsection \ref{sec:UP}, we showed that the unitary $R$ is constructed as either a sequence of rotations $R_{\text{S}}$ or linear combination of unitaries  $R_{\text{LCU}}$ each from the operators in $\mathcal{A}$. Because the size of the anticommuting set is bounded as: $|\mathcal{A}|\leq 2(n-|\mathcal{G}|)+1$ \cite{raussendorf2020phase, kirby2021contextual, ralli2023unitary}, we can use this result to bound the stabilizer rank of the eigenstates of noncontextual Pauli Hamiltonians. We prove the following theorem: 

\begin{theorem}\label{StabRank}
For a general $n$-qubit noncontextual Pauli Hamiltonian, for each eigenvalue there is an associated eigenvector with a stabilizer rank that is linear in the number of qubits. 
\end{theorem}
\textit{Proof}. Given that $\vec{\nu}$ defines a particular eigenstate state $\ket{\psi_\pm(\vec{\nu})}$ (equation \ref{eq:noncon_gs}) with eigenvalue $E_\pm(\vec{\nu})$ (equation \ref{eq:vec_assignment_energy}). Applying $R$ as a linear combination of unitaries ($R_{\text{LCU}}^{\dagger}$, equation \ref{eq:R_LCU}), we observe that there can be at most $|\mathcal{A}| \leq 2(n-|\mathcal{G}|) + 1 \implies \mathcal{O}(2n)$ stabilizer states generated and thus the stabilizer rank of $\ket{\psi_\pm^{nc}(\vec{\nu})} = R_{\text{LCU}}^{\dagger}(\vec{\nu})\ket{\psi_\pm(\vec{\nu})}$ scales linearly with the number of qubits. \qedsymbol{} 

For  $\mathcal{A}\equiv\emptyset$, $R$ must be the identity operator and by equation \ref{eq:noncon_gs} $\ket{\psi_{noncon}(\vec{\nu})}$ will have a stabilizer rank of one.  We thus state the following Corollary:
\begin{Corollary}\label{StabRank2}
For a $n$-qubit noncontextual Pauli Hamiltonian with $\mathcal{A}\equiv\emptyset$, the eigenvectors will have a stabilizer rank of one and $H_{\text{nc}}$ is a Stabilizer Hamiltonian \cite{temme2015fast}.
\end{Corollary}

In summary, the symmetries of the noncontextual Hamiltonian block diagonalize the Hamiltonian. Each assignment of the eigenvalues of the symmetries specifies a stabilizer subspace. The noncommuting operators in $\mathcal{A}$ specify a non-Clifford rotation operator which does not change this stabilizer subspace, but acts nontrivially inside the subspace. This rotation diagonalizes the block, assigning each state to positive and negative eigenspaces. Structure in this non-Clifford operator results in the eigenvectors within each block having stabilizer rank linear in the number of qubits.

The significance of Theorem \ref{StabRank} is that given any eigenvalue of a noncontextual Pauli Hamiltonian $E(\vec{\nu})$, it is always possible to obtain the associated eigenstate efficiently on a classical computer. This is useful, as the expectation value of other observables can be calculated. Furthermore, it offers a new class of classically tractable states that could be used as input for quantum algorithms. Such as the input state in quantum phase estimation \cite{kitaev1995quantum} or the variational quantum eigensolver \cite{peruzzo2014variational}.

\subsection{Dimension of the Eigenspaces of Noncontextual Pauli Hamiltonians \label{sec:multiplicity}}

Utilizing Theorem \ref{AntiCommExpec} and the block form of noncontextual Pauli Hamiltonians (equation \ref{eq:projectors2}) implies the following:

\begin{Corollary}\label{NonconMult}
For a general noncontextual Pauli Hamiltonian generated by the independent set $\mathcal{R} \equiv \mathcal{G} \cup \mathcal{A}$ under the Jordan product, in all cases when $|\mathcal{A}| \geq 2$ the multiplicity of each eigenvalue must be a factor of $2^{\lceil \frac{|\mathcal{A}|-1}{2} \rceil  - 1}$.
\end{Corollary}
Note $|\mathcal{A}| < 2$ implies $|\mathcal{A}| \equiv \emptyset$, the noncontextual Hamiltonian is a stabilizer Hamiltonian. While we can determine the degeneracies caused by $\mathcal{A}$, we note it is possible to have further degeneracies generated by the operators in $\mathcal{G}$. We know that the number of eigenvalues for an $n$ qubit Hermitian operator must be $2^{n}$, Let $m_{\lambda}$ be the number of times a particular eigenvalue $\lambda$ occurs then:
\begin{equation}
\begin{aligned}
    \label{eq:multi2}
     2^{n} = \sum_{\lambda} m_{\lambda}  = \sum_{\lambda} \underbrace{k_{\lambda}}_{\text{due to } \mathcal{G} } \underbrace{\big( 2^{\lceil \frac{|\mathcal{A}|-1}{2} \rceil-1} \big)}_{\text{due to } \mathcal{A} }   \\
     \implies 2^{n+1 - \lceil \frac{|\mathcal{A}|-1}{2} \rceil} = \sum_{\lambda} k_{\lambda} .\end{aligned}
\end{equation}
Here $k_{\lambda}$ denotes the degeneracy of eigenvalue $\lambda$ not caused by the anticommuting blocks $H_{\text{nc}} (\vec{\nu})$ (equation \ref{eq:vec_assignment2}).

In the second line of equation \ref{eq:multi2}, we have divided $2^{n}$ by $2^{\lceil \frac{|\mathcal{A}|-1}{2} \rceil-1}$ thus isolating the dependency of different eigenvalues caused by alternate symmetries including non-$\mathbb{Z}_{2}$ symmetries in the Hamiltonian (i.e. symmetries not in $\overline{\mathcal{G}}$ \footnote{Here the overbar indicates the completion of the set $\mathcal{G}$. In lay terms this means any product of operators in the set $\mathcal{G}$.}). Such symmetries could potentially be used to further simplify any search over the eigenspace, as each symmetry must be conserved by associated solutions. An example where this occurs is for \ce{H2} described in the STO-3G basis (a four qubit problem). Kirby \textit{et al.} in \cite{kirby2019contextuality}  proved that the second quantized molecular Hamiltonian is noncontextual. This Hamiltonian must commute with the number operator, which is a non-$\mathbb{Z}_{2}$ symmetry and thus is an example of where further degeneracies can arise.

\section{Conclusion} \label{sec:conclusion}
In this paper we analyse the properties of noncontextual Pauli Hamiltonians $H_{\text{nc}}$. We first showed how the compatibility graph of noncontextual Pauli Hamiltonians permits a convenient graph interpretation as a disjoint union of commuting cliques (fully connected subgraphs that have no edges between each other) and a set vertices connected with all vertices in the graph.  The commutation structure (Thoerem \ref{NonconDefWill}) ensures that definite eigenvalues can be assigned to all Pauli operators in $H_{\text{nc}}$ without incurring any logical contradictions. This underpins why a noncontextual model can be constructed for such Hermitian operators. 


We then proved that noncontextual Pauli Hamiltonians can have a support of at most $2^{n+1}$ Pauli operators, which spans a larger space than Hamiltonians composed of pairwise commuting Pauli operators, which have a maximum support of $2^{n}$ Pauli operators. 

Next, we analysed the eigenspectrum of noncontextual Pauli Hamiltonians. First, we showed that every noncontextual Pauli Hamiltonian is block diagonal with respect to operators in the set $\mathcal{G}$ (Theorem \ref{block_theorem}). Specifically, each block is defined by a stabilizer subspace that is efficient to define. Each resulting block is written by a linear combination of pairwise anticommuting Pauli operators plus an identity term (or constant shift). Such linear combinations are efficient to diagonalize (Theorem \ref{AntiCommExpec}). This result improves the work in \cite{kirby2020classical}, as it allows noncontextual eigenstates to be directly written down rather than having to perform an optimization over a hypersphere.

We further showed that the unitary required to diagonalize each block in a general $n$-qubit noncontextual Pauli Hamiltonian has specific structure that results in all eigenvalues having an associated eigenvector with stabilizer rank that is linear in the number of qubits. 

The ability to generate a noncontextual eigenstate efficiently via a classical computation offers a benefit to solving noncontextual Pauli Hamiltonians \cite{kirby2020classical} and the CS-VQE algorithm \cite{kirby2021contextual}. It further opens the field to different noncontextual Ansatz that have a linear stabilizer rank. For example in quantum chemistry it is common to start calculations from the approximate Hartree-Fock solution, instead the noncontextual solution could be used to instantiate the problem. We leave such a study to future work. Such initializations would be similar to work that uses a superposition of stabilizer states as input, such as the Clifford Ansatz For Quantum Accuracy (CAFQA) \cite{ravi2022cafqa}, but rather than enforcing a given ansatz to be Clifford (e.g. by rounding the rotation angles in all single qubit rotation gates to integer values of $\frac{\pi}{2}$), the trial state will be motivated by a noncontextual model of the system under consideration. 


We remark that one might also consider the converse to what was shown in this work: given a linear stabilizer rank state does there exist a parent noncontextual Pauli Hamiltonian for which it is an eigenstate? In cases where a quantum state has a noncontextual parent Hamiltonian, it could be useful for adiabatic state preparation \cite{aharonov2003adiabatic, albash2018adiabatic}, given that we can efficiently prepare noncontextual eigenstates on a quantum computer. Given that linear rank stabilizer states are classically efficient to describe, the truth of this statement would imply that the noncontextual problem represents a general notion of classically simulable Hamiltonians. On the other hand, if this proposition does not hold there must therefore exist a broader class of exactly simulable Hamiltonians, which would be an interesting avenue for further research; work related to this idea has been performed in~\cite{patel2023extension, patel2024exactly}.

All work to date, including the present paper is based on Pauli Hamiltonians. Extending the notion of a noncontextual Hamiltonian beyond Pauli Hamiltonians, for example to more general sparse Hamiltonians, is a natural and interesting open question.

\section*{Acknowledgements} \label{sec:acknowledgements}
A.R. and P.J.L. acknowledge support by the NSF STAQ project (PHY-1818914/232580). T.W. also acknowledges support from the Engineering and Physical Sciences Research Council (EP/S021582/1) and CBKSciCon Ltd. 

\bibliographystyle{apsrev4-1.bst}
\bibliography{references.bib}

\begin{thebibliography}{75}%
\makeatletter
\providecommand \@ifxundefined [1]{%
 \@ifx{#1\undefined}
}%
\providecommand \@ifnum [1]{%
 \ifnum #1\expandafter \@firstoftwo
 \else \expandafter \@secondoftwo
 \fi
}%
\providecommand \@ifx [1]{%
 \ifx #1\expandafter \@firstoftwo
 \else \expandafter \@secondoftwo
 \fi
}%
\providecommand \natexlab [1]{#1}%
\providecommand \enquote  [1]{``#1''}%
\providecommand \bibnamefont  [1]{#1}%
\providecommand \bibfnamefont [1]{#1}%
\providecommand \citenamefont [1]{#1}%
\providecommand \href@noop [0]{\@secondoftwo}%
\providecommand \href [0]{\begingroup \@sanitize@url \@href}%
\providecommand \@href[1]{\@@startlink{#1}\@@href}%
\providecommand \@@href[1]{\endgroup#1\@@endlink}%
\providecommand \@sanitize@url [0]{\catcode `\\12\catcode `\$12\catcode
  `\&12\catcode `\#12\catcode `\^12\catcode `\_12\catcode `\%12\relax}%
\providecommand \@@startlink[1]{}%
\providecommand \@@endlink[0]{}%
\providecommand \url  [0]{\begingroup\@sanitize@url \@url }%
\providecommand \@url [1]{\endgroup\@href {#1}{\urlprefix }}%
\providecommand \urlprefix  [0]{URL }%
\providecommand \Eprint [0]{\href }%
\providecommand \doibase [0]{http://dx.doi.org/}%
\providecommand \selectlanguage [0]{\@gobble}%
\providecommand \bibinfo  [0]{\@secondoftwo}%
\providecommand \bibfield  [0]{\@secondoftwo}%
\providecommand \translation [1]{[#1]}%
\providecommand \BibitemOpen [0]{}%
\providecommand \bibitemStop [0]{}%
\providecommand \bibitemNoStop [0]{.\EOS\space}%
\providecommand \EOS [0]{\spacefactor3000\relax}%
\providecommand \BibitemShut  [1]{\csname bibitem#1\endcsname}%
\let\auto@bib@innerbib\@empty
\bibitem [{\citenamefont {Preskill}(2018)}]{preskill2018quantum}%
  \BibitemOpen
  \bibfield  {author} {\bibinfo {author} {\bibfnamefont {J.}~\bibnamefont
  {Preskill}},\ }\href {\doibase https://doi.org/10.22331/q-2018-08-06-79}
  {\bibfield  {journal} {\bibinfo  {journal} {Quantum}\ }\textbf {\bibinfo
  {volume} {2}},\ \bibinfo {pages} {79} (\bibinfo {year} {2018})}\BibitemShut
  {NoStop}%
\bibitem [{\citenamefont {Kirby}\ \emph {et~al.}(2021)\citenamefont {Kirby},
  \citenamefont {Tranter},\ and\ \citenamefont {Love}}]{kirby2021contextual}%
  \BibitemOpen
  \bibfield  {author} {\bibinfo {author} {\bibfnamefont {W.~M.}\ \bibnamefont
  {Kirby}}, \bibinfo {author} {\bibfnamefont {A.}~\bibnamefont {Tranter}}, \
  and\ \bibinfo {author} {\bibfnamefont {P.~J.}\ \bibnamefont {Love}},\ }\href
  {https://quantum-journal.org/papers/q-2021-05-14-456/} {\bibfield  {journal}
  {\bibinfo  {journal} {Quantum}\ }\textbf {\bibinfo {volume} {5}},\ \bibinfo
  {pages} {456} (\bibinfo {year} {2021})}\BibitemShut {NoStop}%
\bibitem [{\citenamefont {Kirby}\ and\ \citenamefont
  {Love}(2019)}]{kirby2019contextuality}%
  \BibitemOpen
  \bibfield  {author} {\bibinfo {author} {\bibfnamefont {W.~M.}\ \bibnamefont
  {Kirby}}\ and\ \bibinfo {author} {\bibfnamefont {P.~J.}\ \bibnamefont
  {Love}},\ }\href@noop {} {\bibfield  {journal} {\bibinfo  {journal} {Physical
  {R}eview {L}etters}\ }\textbf {\bibinfo {volume} {123}},\ \bibinfo {pages}
  {200501} (\bibinfo {year} {2019})}\BibitemShut {NoStop}%
\bibitem [{\citenamefont {Kirby}\ and\ \citenamefont
  {Love}(2020)}]{kirby2020classical}%
  \BibitemOpen
  \bibfield  {author} {\bibinfo {author} {\bibfnamefont {W.~M.}\ \bibnamefont
  {Kirby}}\ and\ \bibinfo {author} {\bibfnamefont {P.~J.}\ \bibnamefont
  {Love}},\ }\href {\doibase 10.1103/PhysRevA.102.032418} {\bibfield  {journal}
  {\bibinfo  {journal} {Physical {R}eview A}\ }\textbf {\bibinfo {volume}
  {102}},\ \bibinfo {pages} {032418} (\bibinfo {year} {2020})},\ \Eprint
  {http://arxiv.org/abs/2002.05693} {arXiv:2002.05693} \BibitemShut {NoStop}%
\bibitem [{\citenamefont {Weaving}\ \emph
  {et~al.}(2023{\natexlab{a}})\citenamefont {Weaving}, \citenamefont {Ralli},
  \citenamefont {Kirby}, \citenamefont {Tranter}, \citenamefont {Love},\ and\
  \citenamefont {Coveney}}]{weaving2023stabilizer}%
  \BibitemOpen
  \bibfield  {author} {\bibinfo {author} {\bibfnamefont {T.}~\bibnamefont
  {Weaving}}, \bibinfo {author} {\bibfnamefont {A.}~\bibnamefont {Ralli}},
  \bibinfo {author} {\bibfnamefont {W.~M.}\ \bibnamefont {Kirby}}, \bibinfo
  {author} {\bibfnamefont {A.}~\bibnamefont {Tranter}}, \bibinfo {author}
  {\bibfnamefont {P.~J.}\ \bibnamefont {Love}}, \ and\ \bibinfo {author}
  {\bibfnamefont {P.~V.}\ \bibnamefont {Coveney}},\ }\href@noop {} {\bibfield
  {journal} {\bibinfo  {journal} {Journal of Chemical Theory and Computation}\
  }\textbf {\bibinfo {volume} {19}},\ \bibinfo {pages} {808} (\bibinfo {year}
  {2023}{\natexlab{a}})}\BibitemShut {NoStop}%
\bibitem [{\citenamefont {Ralli}\ \emph {et~al.}(2023)\citenamefont {Ralli},
  \citenamefont {Weaving}, \citenamefont {Tranter}, \citenamefont {Kirby},
  \citenamefont {Love},\ and\ \citenamefont {Coveney}}]{ralli2023unitary}%
  \BibitemOpen
  \bibfield  {author} {\bibinfo {author} {\bibfnamefont {A.}~\bibnamefont
  {Ralli}}, \bibinfo {author} {\bibfnamefont {T.}~\bibnamefont {Weaving}},
  \bibinfo {author} {\bibfnamefont {A.}~\bibnamefont {Tranter}}, \bibinfo
  {author} {\bibfnamefont {W.~M.}\ \bibnamefont {Kirby}}, \bibinfo {author}
  {\bibfnamefont {P.~J.}\ \bibnamefont {Love}}, \ and\ \bibinfo {author}
  {\bibfnamefont {P.~V.}\ \bibnamefont {Coveney}},\ }\href@noop {} {\bibfield
  {journal} {\bibinfo  {journal} {Physical {R}eview Research}\ }\textbf
  {\bibinfo {volume} {5}},\ \bibinfo {pages} {013095} (\bibinfo {year}
  {2023})}\BibitemShut {NoStop}%
\bibitem [{\citenamefont {Weaving}\ \emph
  {et~al.}(2023{\natexlab{b}})\citenamefont {Weaving}, \citenamefont {Ralli},
  \citenamefont {Kirby}, \citenamefont {Love}, \citenamefont {Succi},\ and\
  \citenamefont {Coveney}}]{weaving2023benchmarking}%
  \BibitemOpen
  \bibfield  {author} {\bibinfo {author} {\bibfnamefont {T.}~\bibnamefont
  {Weaving}}, \bibinfo {author} {\bibfnamefont {A.}~\bibnamefont {Ralli}},
  \bibinfo {author} {\bibfnamefont {W.~M.}\ \bibnamefont {Kirby}}, \bibinfo
  {author} {\bibfnamefont {P.~J.}\ \bibnamefont {Love}}, \bibinfo {author}
  {\bibfnamefont {S.}~\bibnamefont {Succi}}, \ and\ \bibinfo {author}
  {\bibfnamefont {P.~V.}\ \bibnamefont {Coveney}},\ }\href {\doibase
  https://doi.org/10.1103/PhysRevResearch.5.043054} {\bibfield  {journal}
  {\bibinfo  {journal} {Physical Review Research}\ }\textbf {\bibinfo {volume}
  {5}},\ \bibinfo {pages} {043054} (\bibinfo {year}
  {2023}{\natexlab{b}})}\BibitemShut {NoStop}%
\bibitem [{\citenamefont {Weaving}\ \emph {et~al.}(2025)\citenamefont
  {Weaving}, \citenamefont {Ralli}, \citenamefont {Love}, \citenamefont
  {Succi},\ and\ \citenamefont {Coveney}}]{weaving2025contextual}%
  \BibitemOpen
  \bibfield  {author} {\bibinfo {author} {\bibfnamefont {T.}~\bibnamefont
  {Weaving}}, \bibinfo {author} {\bibfnamefont {A.}~\bibnamefont {Ralli}},
  \bibinfo {author} {\bibfnamefont {P.~J.}\ \bibnamefont {Love}}, \bibinfo
  {author} {\bibfnamefont {S.}~\bibnamefont {Succi}}, \ and\ \bibinfo {author}
  {\bibfnamefont {P.~V.}\ \bibnamefont {Coveney}},\ }\href {\doibase
  https://doi.org/10.1038/s41534-024-00952-4} {\bibfield  {journal} {\bibinfo
  {journal} {npj Quantum Information}\ }\textbf {\bibinfo {volume} {11}},\
  \bibinfo {pages} {25} (\bibinfo {year} {2025})}\BibitemShut {NoStop}%
\bibitem [{\citenamefont {Preskill}(2012)}]{preskill2012quantum}%
  \BibitemOpen
  \bibfield  {author} {\bibinfo {author} {\bibfnamefont {J.}~\bibnamefont
  {Preskill}},\ }\href@noop {} {\bibfield  {journal} {\bibinfo  {journal}
  {arXiv preprint arXiv:1203.5813}\ } (\bibinfo {year} {2012})}\BibitemShut
  {NoStop}%
\bibitem [{\citenamefont {Howard}\ \emph {et~al.}(2014)\citenamefont {Howard},
  \citenamefont {Wallman}, \citenamefont {Veitch},\ and\ \citenamefont
  {Emerson}}]{howard2014contextuality}%
  \BibitemOpen
  \bibfield  {author} {\bibinfo {author} {\bibfnamefont {M.}~\bibnamefont
  {Howard}}, \bibinfo {author} {\bibfnamefont {J.}~\bibnamefont {Wallman}},
  \bibinfo {author} {\bibfnamefont {V.}~\bibnamefont {Veitch}}, \ and\ \bibinfo
  {author} {\bibfnamefont {J.}~\bibnamefont {Emerson}},\ }\href@noop {}
  {\bibfield  {journal} {\bibinfo  {journal} {Nature}\ }\textbf {\bibinfo
  {volume} {510}},\ \bibinfo {pages} {351} (\bibinfo {year}
  {2014})}\BibitemShut {NoStop}%
\bibitem [{\citenamefont {Shahandeh}(2021)}]{shahandeh2021quantum}%
  \BibitemOpen
  \bibfield  {author} {\bibinfo {author} {\bibfnamefont {F.}~\bibnamefont
  {Shahandeh}},\ }\href@noop {} {\bibfield  {journal} {\bibinfo  {journal}
  {arXiv preprint arXiv:2112.00024}\ } (\bibinfo {year} {2021})}\BibitemShut
  {NoStop}%
\bibitem [{\citenamefont {Bermejo-Vega}\ \emph {et~al.}(2017)\citenamefont
  {Bermejo-Vega}, \citenamefont {Delfosse}, \citenamefont {Browne},
  \citenamefont {Okay},\ and\ \citenamefont
  {Raussendorf}}]{bermejo2017contextuality}%
  \BibitemOpen
  \bibfield  {author} {\bibinfo {author} {\bibfnamefont {J.}~\bibnamefont
  {Bermejo-Vega}}, \bibinfo {author} {\bibfnamefont {N.}~\bibnamefont
  {Delfosse}}, \bibinfo {author} {\bibfnamefont {D.~E.}\ \bibnamefont
  {Browne}}, \bibinfo {author} {\bibfnamefont {C.}~\bibnamefont {Okay}}, \ and\
  \bibinfo {author} {\bibfnamefont {R.}~\bibnamefont {Raussendorf}},\
  }\href@noop {} {\bibfield  {journal} {\bibinfo  {journal} {Physical {R}eview
  {L}etters}\ }\textbf {\bibinfo {volume} {119}},\ \bibinfo {pages} {120505}
  (\bibinfo {year} {2017})}\BibitemShut {NoStop}%
\bibitem [{\citenamefont {Raussendorf}(2013)}]{raussendorf2013contextuality}%
  \BibitemOpen
  \bibfield  {author} {\bibinfo {author} {\bibfnamefont {R.}~\bibnamefont
  {Raussendorf}},\ }\href@noop {} {\bibfield  {journal} {\bibinfo  {journal}
  {Physical {R}eview A}\ }\textbf {\bibinfo {volume} {88}},\ \bibinfo {pages}
  {022322} (\bibinfo {year} {2013})}\BibitemShut {NoStop}%
\bibitem [{\citenamefont {Delfosse}\ \emph {et~al.}(2015)\citenamefont
  {Delfosse}, \citenamefont {Guerin}, \citenamefont {Bian},\ and\ \citenamefont
  {Raussendorf}}]{delfosse2015wigner}%
  \BibitemOpen
  \bibfield  {author} {\bibinfo {author} {\bibfnamefont {N.}~\bibnamefont
  {Delfosse}}, \bibinfo {author} {\bibfnamefont {P.~A.}\ \bibnamefont
  {Guerin}}, \bibinfo {author} {\bibfnamefont {J.}~\bibnamefont {Bian}}, \ and\
  \bibinfo {author} {\bibfnamefont {R.}~\bibnamefont {Raussendorf}},\
  }\href@noop {} {\bibfield  {journal} {\bibinfo  {journal} {Physical {R}eview
  X}\ }\textbf {\bibinfo {volume} {5}},\ \bibinfo {pages} {021003} (\bibinfo
  {year} {2015})}\BibitemShut {NoStop}%
\bibitem [{\citenamefont {Mansfield}\ and\ \citenamefont
  {Kashefi}(2018)}]{mansfield2018quantum}%
  \BibitemOpen
  \bibfield  {author} {\bibinfo {author} {\bibfnamefont {S.}~\bibnamefont
  {Mansfield}}\ and\ \bibinfo {author} {\bibfnamefont {E.}~\bibnamefont
  {Kashefi}},\ }\href@noop {} {\bibfield  {journal} {\bibinfo  {journal}
  {Physical~{R}eview~{Letters}}\ }\textbf {\bibinfo {volume} {121}},\ \bibinfo
  {pages} {230401} (\bibinfo {year} {2018})}\BibitemShut {NoStop}%
\bibitem [{\citenamefont {Karanjai}\ \emph {et~al.}(2018)\citenamefont
  {Karanjai}, \citenamefont {Wallman},\ and\ \citenamefont
  {Bartlett}}]{karanjai2018contextuality}%
  \BibitemOpen
  \bibfield  {author} {\bibinfo {author} {\bibfnamefont {A.}~\bibnamefont
  {Karanjai}}, \bibinfo {author} {\bibfnamefont {J.~J.}\ \bibnamefont
  {Wallman}}, \ and\ \bibinfo {author} {\bibfnamefont {S.~D.}\ \bibnamefont
  {Bartlett}},\ }\href@noop {} {\bibfield  {journal} {\bibinfo  {journal}
  {arXiv preprint arXiv:1802.07744}\ } (\bibinfo {year} {2018})}\BibitemShut
  {NoStop}%
\bibitem [{\citenamefont {Grudka}\ \emph {et~al.}(2014)\citenamefont {Grudka},
  \citenamefont {Horodecki}, \citenamefont {Horodecki}, \citenamefont
  {Horodecki}, \citenamefont {Horodecki}, \citenamefont {Joshi}, \citenamefont
  {K{\l}obus},\ and\ \citenamefont {W{\'o}jcik}}]{grudka2014quantifying}%
  \BibitemOpen
  \bibfield  {author} {\bibinfo {author} {\bibfnamefont {A.}~\bibnamefont
  {Grudka}}, \bibinfo {author} {\bibfnamefont {K.}~\bibnamefont {Horodecki}},
  \bibinfo {author} {\bibfnamefont {M.}~\bibnamefont {Horodecki}}, \bibinfo
  {author} {\bibfnamefont {P.}~\bibnamefont {Horodecki}}, \bibinfo {author}
  {\bibfnamefont {R.}~\bibnamefont {Horodecki}}, \bibinfo {author}
  {\bibfnamefont {P.}~\bibnamefont {Joshi}}, \bibinfo {author} {\bibfnamefont
  {W.}~\bibnamefont {K{\l}obus}}, \ and\ \bibinfo {author} {\bibfnamefont
  {A.}~\bibnamefont {W{\'o}jcik}},\ }\href@noop {} {\bibfield  {journal}
  {\bibinfo  {journal} {Physical {R}eview {L}etters}\ }\textbf {\bibinfo
  {volume} {112}},\ \bibinfo {pages} {120401} (\bibinfo {year}
  {2014})}\BibitemShut {NoStop}%
\bibitem [{\citenamefont {Budroni}\ \emph {et~al.}(2022)\citenamefont
  {Budroni}, \citenamefont {Cabello}, \citenamefont {G{\"u}hne}, \citenamefont
  {Kleinmann},\ and\ \citenamefont {Larsson}}]{budroni2022kochen}%
  \BibitemOpen
  \bibfield  {author} {\bibinfo {author} {\bibfnamefont {C.}~\bibnamefont
  {Budroni}}, \bibinfo {author} {\bibfnamefont {A.}~\bibnamefont {Cabello}},
  \bibinfo {author} {\bibfnamefont {O.}~\bibnamefont {G{\"u}hne}}, \bibinfo
  {author} {\bibfnamefont {M.}~\bibnamefont {Kleinmann}}, \ and\ \bibinfo
  {author} {\bibfnamefont {J.-{\AA}.}\ \bibnamefont {Larsson}},\ }\href@noop {}
  {\bibfield  {journal} {\bibinfo  {journal} {{R}eviews of Modern Physics}\
  }\textbf {\bibinfo {volume} {94}},\ \bibinfo {pages} {045007} (\bibinfo
  {year} {2022})}\BibitemShut {NoStop}%
\bibitem [{\citenamefont {Gupta}\ \emph {et~al.}(2023)\citenamefont {Gupta},
  \citenamefont {Saha}, \citenamefont {Xu}, \citenamefont {Cabello},\ and\
  \citenamefont {Majumdar}}]{gupta2023quantum}%
  \BibitemOpen
  \bibfield  {author} {\bibinfo {author} {\bibfnamefont {S.}~\bibnamefont
  {Gupta}}, \bibinfo {author} {\bibfnamefont {D.}~\bibnamefont {Saha}},
  \bibinfo {author} {\bibfnamefont {Z.-P.}\ \bibnamefont {Xu}}, \bibinfo
  {author} {\bibfnamefont {A.}~\bibnamefont {Cabello}}, \ and\ \bibinfo
  {author} {\bibfnamefont {A.}~\bibnamefont {Majumdar}},\ }\href@noop {}
  {\bibfield  {journal} {\bibinfo  {journal} {Physical {R}eview {L}etters}\
  }\textbf {\bibinfo {volume} {130}},\ \bibinfo {pages} {080802} (\bibinfo
  {year} {2023})}\BibitemShut {NoStop}%
\bibitem [{\citenamefont {Pavicic}(2023)}]{pavicic2023quantum}%
  \BibitemOpen
  \bibfield  {author} {\bibinfo {author} {\bibfnamefont {M.}~\bibnamefont
  {Pavicic}},\ }\href@noop {} {\bibfield  {journal} {\bibinfo  {journal}
  {Quantum}\ }\textbf {\bibinfo {volume} {7}},\ \bibinfo {pages} {953}
  (\bibinfo {year} {2023})}\BibitemShut {NoStop}%
\bibitem [{\citenamefont {Spekkens}(2008)}]{spekkens2008negativity}%
  \BibitemOpen
  \bibfield  {author} {\bibinfo {author} {\bibfnamefont {R.~W.}\ \bibnamefont
  {Spekkens}},\ }\href@noop {} {\bibfield  {journal} {\bibinfo  {journal}
  {Physical review letters}\ }\textbf {\bibinfo {volume} {101}},\ \bibinfo
  {pages} {020401} (\bibinfo {year} {2008})}\BibitemShut {NoStop}%
\bibitem [{\citenamefont {Ferrie}\ and\ \citenamefont
  {Emerson}(2008)}]{ferrie2008frame}%
  \BibitemOpen
  \bibfield  {author} {\bibinfo {author} {\bibfnamefont {C.}~\bibnamefont
  {Ferrie}}\ and\ \bibinfo {author} {\bibfnamefont {J.}~\bibnamefont
  {Emerson}},\ }\href@noop {} {\bibfield  {journal} {\bibinfo  {journal}
  {Journal of Physics A: Mathematical and Theoretical}\ }\textbf {\bibinfo
  {volume} {41}},\ \bibinfo {pages} {352001} (\bibinfo {year}
  {2008})}\BibitemShut {NoStop}%
\bibitem [{\citenamefont {Ferrie}\ and\ \citenamefont
  {Emerson}(2009)}]{ferrie2009framed}%
  \BibitemOpen
  \bibfield  {author} {\bibinfo {author} {\bibfnamefont {C.}~\bibnamefont
  {Ferrie}}\ and\ \bibinfo {author} {\bibfnamefont {J.}~\bibnamefont
  {Emerson}},\ }\href@noop {} {\bibfield  {journal} {\bibinfo  {journal} {New
  Journal of Physics}\ }\textbf {\bibinfo {volume} {11}},\ \bibinfo {pages}
  {063040} (\bibinfo {year} {2009})}\BibitemShut {NoStop}%
\bibitem [{\citenamefont {Kocia}\ \emph
  {et~al.}(2017{\natexlab{a}})\citenamefont {Kocia}, \citenamefont {Huang},\
  and\ \citenamefont {Love}}]{kocia2017semiclassical}%
  \BibitemOpen
  \bibfield  {author} {\bibinfo {author} {\bibfnamefont {L.}~\bibnamefont
  {Kocia}}, \bibinfo {author} {\bibfnamefont {Y.}~\bibnamefont {Huang}}, \ and\
  \bibinfo {author} {\bibfnamefont {P.}~\bibnamefont {Love}},\ }\href@noop {}
  {\bibfield  {journal} {\bibinfo  {journal} {Physical Review A}\ }\textbf
  {\bibinfo {volume} {96}},\ \bibinfo {pages} {032331} (\bibinfo {year}
  {2017}{\natexlab{a}})}\BibitemShut {NoStop}%
\bibitem [{\citenamefont {Kocia}\ \emph
  {et~al.}(2017{\natexlab{b}})\citenamefont {Kocia}, \citenamefont {Huang},\
  and\ \citenamefont {Love}}]{kocia2017discrete}%
  \BibitemOpen
  \bibfield  {author} {\bibinfo {author} {\bibfnamefont {L.}~\bibnamefont
  {Kocia}}, \bibinfo {author} {\bibfnamefont {Y.}~\bibnamefont {Huang}}, \ and\
  \bibinfo {author} {\bibfnamefont {P.}~\bibnamefont {Love}},\ }\href@noop {}
  {\bibfield  {journal} {\bibinfo  {journal} {Entropy}\ }\textbf {\bibinfo
  {volume} {19}},\ \bibinfo {pages} {353} (\bibinfo {year}
  {2017}{\natexlab{b}})}\BibitemShut {NoStop}%
\bibitem [{\citenamefont {Kocia}\ and\ \citenamefont
  {Love}(2017)}]{kocia2017discrete2}%
  \BibitemOpen
  \bibfield  {author} {\bibinfo {author} {\bibfnamefont {L.}~\bibnamefont
  {Kocia}}\ and\ \bibinfo {author} {\bibfnamefont {P.}~\bibnamefont {Love}},\
  }\href@noop {} {\bibfield  {journal} {\bibinfo  {journal} {Physical Review
  A}\ }\textbf {\bibinfo {volume} {96}},\ \bibinfo {pages} {062134} (\bibinfo
  {year} {2017})}\BibitemShut {NoStop}%
\bibitem [{\citenamefont {Kocia}\ and\ \citenamefont
  {Love}(2018)}]{kocia2018measurement}%
  \BibitemOpen
  \bibfield  {author} {\bibinfo {author} {\bibfnamefont {L.}~\bibnamefont
  {Kocia}}\ and\ \bibinfo {author} {\bibfnamefont {P.}~\bibnamefont {Love}},\
  }\href@noop {} {\bibfield  {journal} {\bibinfo  {journal} {New Journal of
  Physics}\ }\textbf {\bibinfo {volume} {20}},\ \bibinfo {pages} {073020}
  (\bibinfo {year} {2018})}\BibitemShut {NoStop}%
\bibitem [{\citenamefont {Kocia}\ and\ \citenamefont
  {Love}(2019)}]{kocia2019non}%
  \BibitemOpen
  \bibfield  {author} {\bibinfo {author} {\bibfnamefont {L.}~\bibnamefont
  {Kocia}}\ and\ \bibinfo {author} {\bibfnamefont {P.}~\bibnamefont {Love}},\
  }\href@noop {} {\bibfield  {journal} {\bibinfo  {journal} {Journal of Physics
  A: Mathematical and Theoretical}\ }\textbf {\bibinfo {volume} {52}},\
  \bibinfo {pages} {095303} (\bibinfo {year} {2019})}\BibitemShut {NoStop}%
\bibitem [{\citenamefont {Kocia}\ and\ \citenamefont
  {Love}(2021)}]{kocia2021stationary}%
  \BibitemOpen
  \bibfield  {author} {\bibinfo {author} {\bibfnamefont {L.}~\bibnamefont
  {Kocia}}\ and\ \bibinfo {author} {\bibfnamefont {P.}~\bibnamefont {Love}},\
  }\href@noop {} {\bibfield  {journal} {\bibinfo  {journal} {Quantum}\ }\textbf
  {\bibinfo {volume} {5}},\ \bibinfo {pages} {494} (\bibinfo {year}
  {2021})}\BibitemShut {NoStop}%
\bibitem [{\citenamefont {Gottesman}(1996)}]{gottesman1996class}%
  \BibitemOpen
  \bibfield  {author} {\bibinfo {author} {\bibfnamefont {D.}~\bibnamefont
  {Gottesman}},\ }\href@noop {} {\bibfield  {journal} {\bibinfo  {journal}
  {Physical {R}eview A}\ }\textbf {\bibinfo {volume} {54}},\ \bibinfo {pages}
  {1862} (\bibinfo {year} {1996})}\BibitemShut {NoStop}%
\bibitem [{\citenamefont {Gottesman}(1997)}]{gottesman1997stabilizer}%
  \BibitemOpen
  \bibfield  {author} {\bibinfo {author} {\bibfnamefont {D.}~\bibnamefont
  {Gottesman}},\ }\href@noop {} {\emph {\bibinfo {title} {Stabilizer codes and
  quantum error correction}}}\ (\bibinfo  {publisher} {California Institute of
  Technology},\ \bibinfo {year} {1997})\BibitemShut {NoStop}%
\bibitem [{\citenamefont {Gottesman}(1998)}]{gottesman1998heisenberg}%
  \BibitemOpen
  \bibfield  {author} {\bibinfo {author} {\bibfnamefont {D.}~\bibnamefont
  {Gottesman}},\ }\href@noop {} {\bibfield  {journal} {\bibinfo  {journal}
  {arXiv preprint quant-ph/9807006}\ } (\bibinfo {year} {1998})}\BibitemShut
  {NoStop}%
\bibitem [{\citenamefont {Aaronson}\ and\ \citenamefont
  {Gottesman}(2004)}]{aaronson2004improved}%
  \BibitemOpen
  \bibfield  {author} {\bibinfo {author} {\bibfnamefont {S.}~\bibnamefont
  {Aaronson}}\ and\ \bibinfo {author} {\bibfnamefont {D.}~\bibnamefont
  {Gottesman}},\ }\href@noop {} {\bibfield  {journal} {\bibinfo  {journal}
  {Physical {R}eview A}\ }\textbf {\bibinfo {volume} {70}},\ \bibinfo {pages}
  {052328} (\bibinfo {year} {2004})}\BibitemShut {NoStop}%
\bibitem [{\citenamefont {Spekkens}(2014)}]{spekkens2016quasi}%
  \BibitemOpen
  \bibfield  {author} {\bibinfo {author} {\bibfnamefont {R.~W.}\ \bibnamefont
  {Spekkens}},\ }\href {\doibase 10.1007/978-94-017-7303-4} {\bibfield
  {journal} {\bibinfo  {journal} {Fundamental Theories of Physics}\ }\textbf
  {\bibinfo {volume} {181}},\ \bibinfo {pages} {83} (\bibinfo {year} {2014})},\
  \Eprint {http://arxiv.org/abs/1409.5041} {arXiv:1409.5041} \BibitemShut
  {NoStop}%
\bibitem [{\citenamefont {Garcia}\ \emph {et~al.}(2012)\citenamefont {Garcia},
  \citenamefont {Markov},\ and\ \citenamefont {Cross}}]{garcia2012efficient}%
  \BibitemOpen
  \bibfield  {author} {\bibinfo {author} {\bibfnamefont {H.~J.}\ \bibnamefont
  {Garcia}}, \bibinfo {author} {\bibfnamefont {I.~L.}\ \bibnamefont {Markov}},
  \ and\ \bibinfo {author} {\bibfnamefont {A.~W.}\ \bibnamefont {Cross}},\
  }\href@noop {} {\bibfield  {journal} {\bibinfo  {journal} {arXiv preprint
  arXiv:1210.6646}\ } (\bibinfo {year} {2012})}\BibitemShut {NoStop}%
\bibitem [{\citenamefont {Garc{\'\i}a}\ \emph {et~al.}(2014)\citenamefont
  {Garc{\'\i}a}, \citenamefont {Markov},\ and\ \citenamefont
  {Cross}}]{garcia2014geometry}%
  \BibitemOpen
  \bibfield  {author} {\bibinfo {author} {\bibfnamefont {H.~J.}\ \bibnamefont
  {Garc{\'\i}a}}, \bibinfo {author} {\bibfnamefont {I.~L.}\ \bibnamefont
  {Markov}}, \ and\ \bibinfo {author} {\bibfnamefont {A.~W.}\ \bibnamefont
  {Cross}},\ }\href@noop {} {\bibfield  {journal} {\bibinfo  {journal} {Quantum
  Information \& Computation}\ }\textbf {\bibinfo {volume} {14}},\ \bibinfo
  {pages} {683} (\bibinfo {year} {2014})}\BibitemShut {NoStop}%
\bibitem [{\citenamefont {Bravyi}\ \emph {et~al.}(2016)\citenamefont {Bravyi},
  \citenamefont {Smith},\ and\ \citenamefont {Smolin}}]{bravyi2016trading}%
  \BibitemOpen
  \bibfield  {author} {\bibinfo {author} {\bibfnamefont {S.}~\bibnamefont
  {Bravyi}}, \bibinfo {author} {\bibfnamefont {G.}~\bibnamefont {Smith}}, \
  and\ \bibinfo {author} {\bibfnamefont {J.~A.}\ \bibnamefont {Smolin}},\
  }\href@noop {} {\bibfield  {journal} {\bibinfo  {journal} {Physical {R}eview
  X}\ }\textbf {\bibinfo {volume} {6}},\ \bibinfo {pages} {021043} (\bibinfo
  {year} {2016})}\BibitemShut {NoStop}%
\bibitem [{\citenamefont {Bravyi}\ \emph {et~al.}(2019)\citenamefont {Bravyi},
  \citenamefont {Browne}, \citenamefont {Calpin}, \citenamefont {Campbell},
  \citenamefont {Gosset},\ and\ \citenamefont {Howard}}]{bravyi2019simulation}%
  \BibitemOpen
  \bibfield  {author} {\bibinfo {author} {\bibfnamefont {S.}~\bibnamefont
  {Bravyi}}, \bibinfo {author} {\bibfnamefont {D.}~\bibnamefont {Browne}},
  \bibinfo {author} {\bibfnamefont {P.}~\bibnamefont {Calpin}}, \bibinfo
  {author} {\bibfnamefont {E.}~\bibnamefont {Campbell}}, \bibinfo {author}
  {\bibfnamefont {D.}~\bibnamefont {Gosset}}, \ and\ \bibinfo {author}
  {\bibfnamefont {M.}~\bibnamefont {Howard}},\ }\href@noop {} {\bibfield
  {journal} {\bibinfo  {journal} {Quantum}\ }\textbf {\bibinfo {volume} {3}},\
  \bibinfo {pages} {181} (\bibinfo {year} {2019})}\BibitemShut {NoStop}%
\bibitem [{\citenamefont {Bravyi}\ and\ \citenamefont
  {Gosset}(2016)}]{PhysRevLett.116.250501}%
  \BibitemOpen
  \bibfield  {author} {\bibinfo {author} {\bibfnamefont {S.}~\bibnamefont
  {Bravyi}}\ and\ \bibinfo {author} {\bibfnamefont {D.}~\bibnamefont
  {Gosset}},\ }\href {\doibase 10.1103/PhysRevLett.116.250501} {\bibfield
  {journal} {\bibinfo  {journal} {Phys. Rev. Lett.}\ }\textbf {\bibinfo
  {volume} {116}},\ \bibinfo {pages} {250501} (\bibinfo {year}
  {2016})}\BibitemShut {NoStop}%
\bibitem [{\citenamefont {Kitaev}(2003)}]{kitaev2003fault}%
  \BibitemOpen
  \bibfield  {author} {\bibinfo {author} {\bibfnamefont {A.~Y.}\ \bibnamefont
  {Kitaev}},\ }\href@noop {} {\bibfield  {journal} {\bibinfo  {journal} {Annals
  of physics}\ }\textbf {\bibinfo {volume} {303}},\ \bibinfo {pages} {2}
  (\bibinfo {year} {2003})}\BibitemShut {NoStop}%
\bibitem [{\citenamefont {Dennis}\ \emph {et~al.}(2002)\citenamefont {Dennis},
  \citenamefont {Kitaev}, \citenamefont {Landahl},\ and\ \citenamefont
  {Preskill}}]{dennis2002topological}%
  \BibitemOpen
  \bibfield  {author} {\bibinfo {author} {\bibfnamefont {E.}~\bibnamefont
  {Dennis}}, \bibinfo {author} {\bibfnamefont {A.}~\bibnamefont {Kitaev}},
  \bibinfo {author} {\bibfnamefont {A.}~\bibnamefont {Landahl}}, \ and\
  \bibinfo {author} {\bibfnamefont {J.}~\bibnamefont {Preskill}},\ }\href@noop
  {} {\bibfield  {journal} {\bibinfo  {journal} {Journal of Mathematical
  Physics}\ }\textbf {\bibinfo {volume} {43}},\ \bibinfo {pages} {4452}
  (\bibinfo {year} {2002})}\BibitemShut {NoStop}%
\bibitem [{\citenamefont {Bravyi}\ and\ \citenamefont
  {Gosset}(2015)}]{bravyi2015gapped}%
  \BibitemOpen
  \bibfield  {author} {\bibinfo {author} {\bibfnamefont {S.}~\bibnamefont
  {Bravyi}}\ and\ \bibinfo {author} {\bibfnamefont {D.}~\bibnamefont
  {Gosset}},\ }\href@noop {} {\bibfield  {journal} {\bibinfo  {journal}
  {Journal of Mathematical Physics}\ }\textbf {\bibinfo {volume} {56}}
  (\bibinfo {year} {2015})}\BibitemShut {NoStop}%
\bibitem [{\citenamefont {Bravyi}\ \emph {et~al.}(2006)\citenamefont {Bravyi},
  \citenamefont {Divincenzo}, \citenamefont {Oliveira},\ and\ \citenamefont
  {Terhal}}]{bravyi2006complexity}%
  \BibitemOpen
  \bibfield  {author} {\bibinfo {author} {\bibfnamefont {S.}~\bibnamefont
  {Bravyi}}, \bibinfo {author} {\bibfnamefont {D.~P.}\ \bibnamefont
  {Divincenzo}}, \bibinfo {author} {\bibfnamefont {R.~I.}\ \bibnamefont
  {Oliveira}}, \ and\ \bibinfo {author} {\bibfnamefont {B.~M.}\ \bibnamefont
  {Terhal}},\ }\href@noop {} {\bibfield  {journal} {\bibinfo  {journal} {arXiv
  preprint quant-ph/0606140}\ } (\bibinfo {year} {2006})}\BibitemShut {NoStop}%
\bibitem [{\citenamefont {Kitaev}\ \emph {et~al.}(2002)\citenamefont {Kitaev},
  \citenamefont {Shen},\ and\ \citenamefont {Vyalyi}}]{kitaev2002classical}%
  \BibitemOpen
  \bibfield  {author} {\bibinfo {author} {\bibfnamefont {A.~Y.}\ \bibnamefont
  {Kitaev}}, \bibinfo {author} {\bibfnamefont {A.}~\bibnamefont {Shen}}, \ and\
  \bibinfo {author} {\bibfnamefont {M.~N.}\ \bibnamefont {Vyalyi}},\
  }\href@noop {} {\emph {\bibinfo {title} {Classical and quantum
  computation}}},\ \bibinfo {number} {47}\ (\bibinfo  {publisher} {American
  Mathematical Soc.},\ \bibinfo {year} {2002})\BibitemShut {NoStop}%
\bibitem [{\citenamefont {Kempe}\ \emph {et~al.}(2006)\citenamefont {Kempe},
  \citenamefont {Kitaev},\ and\ \citenamefont {Regev}}]{kempe2006complexity}%
  \BibitemOpen
  \bibfield  {author} {\bibinfo {author} {\bibfnamefont {J.}~\bibnamefont
  {Kempe}}, \bibinfo {author} {\bibfnamefont {A.}~\bibnamefont {Kitaev}}, \
  and\ \bibinfo {author} {\bibfnamefont {O.}~\bibnamefont {Regev}},\
  }\href@noop {} {\bibfield  {journal} {\bibinfo  {journal} {Siam journal on
  computing}\ }\textbf {\bibinfo {volume} {35}},\ \bibinfo {pages} {1070}
  (\bibinfo {year} {2006})}\BibitemShut {NoStop}%
\bibitem [{\citenamefont {Cubitt}\ and\ \citenamefont
  {Montanaro}(2016)}]{cubitt2016complexity}%
  \BibitemOpen
  \bibfield  {author} {\bibinfo {author} {\bibfnamefont {T.}~\bibnamefont
  {Cubitt}}\ and\ \bibinfo {author} {\bibfnamefont {A.}~\bibnamefont
  {Montanaro}},\ }\href@noop {} {\bibfield  {journal} {\bibinfo  {journal}
  {SIAM Journal on Computing}\ }\textbf {\bibinfo {volume} {45}},\ \bibinfo
  {pages} {268} (\bibinfo {year} {2016})}\BibitemShut {NoStop}%
\bibitem [{\citenamefont {Osborne}(2012)}]{osborne2012hamiltonian}%
  \BibitemOpen
  \bibfield  {author} {\bibinfo {author} {\bibfnamefont {T.~J.}\ \bibnamefont
  {Osborne}},\ }\href@noop {} {\bibfield  {journal} {\bibinfo  {journal}
  {Reports on progress in physics}\ }\textbf {\bibinfo {volume} {75}},\
  \bibinfo {pages} {022001} (\bibinfo {year} {2012})}\BibitemShut {NoStop}%
\bibitem [{\citenamefont {Bravyi}\ and\ \citenamefont
  {Vyalyi}(2003)}]{bravyi2003commutative}%
  \BibitemOpen
  \bibfield  {author} {\bibinfo {author} {\bibfnamefont {S.}~\bibnamefont
  {Bravyi}}\ and\ \bibinfo {author} {\bibfnamefont {M.}~\bibnamefont
  {Vyalyi}},\ }\href@noop {} {\bibfield  {journal} {\bibinfo  {journal} {arXiv
  preprint quant-ph/0308021}\ } (\bibinfo {year} {2003})}\BibitemShut {NoStop}%
\bibitem [{\citenamefont {Bravyi}(2006)}]{bravyi2006efficient}%
  \BibitemOpen
  \bibfield  {author} {\bibinfo {author} {\bibfnamefont {S.}~\bibnamefont
  {Bravyi}},\ }\href@noop {} {\bibfield  {journal} {\bibinfo  {journal} {arXiv
  preprint quant-ph/0602108}\ } (\bibinfo {year} {2006})}\BibitemShut {NoStop}%
\bibitem [{\citenamefont {Bravyi}\ and\ \citenamefont
  {Terhal}(2010)}]{bravyi2010complexity}%
  \BibitemOpen
  \bibfield  {author} {\bibinfo {author} {\bibfnamefont {S.}~\bibnamefont
  {Bravyi}}\ and\ \bibinfo {author} {\bibfnamefont {B.}~\bibnamefont
  {Terhal}},\ }\href@noop {} {\bibfield  {journal} {\bibinfo  {journal} {Siam
  journal on computing}\ }\textbf {\bibinfo {volume} {39}},\ \bibinfo {pages}
  {1462} (\bibinfo {year} {2010})}\BibitemShut {NoStop}%
\bibitem [{\citenamefont {Aharonov}\ and\ \citenamefont
  {Eldar}(2011)}]{aharonov2011complexity}%
  \BibitemOpen
  \bibfield  {author} {\bibinfo {author} {\bibfnamefont {D.}~\bibnamefont
  {Aharonov}}\ and\ \bibinfo {author} {\bibfnamefont {L.}~\bibnamefont
  {Eldar}},\ }in\ \href@noop {} {\emph {\bibinfo {booktitle} {2011 IEEE 52nd
  Annual Symposium on Foundations of Computer Science}}}\ (\bibinfo
  {organization} {IEEE},\ \bibinfo {year} {2011})\ pp.\ \bibinfo {pages}
  {334--343}\BibitemShut {NoStop}%
\bibitem [{\citenamefont {Hofer-Szab{\'o}}(2022)}]{hofer2022two}%
  \BibitemOpen
  \bibfield  {author} {\bibinfo {author} {\bibfnamefont {G.}~\bibnamefont
  {Hofer-Szab{\'o}}},\ }\href {\doibase
  https://doi.org/10.1016/j.shpsa.2022.02.012} {\bibfield  {journal} {\bibinfo
  {journal} {Studies in History and Philosophy of Science}\ }\textbf {\bibinfo
  {volume} {93}},\ \bibinfo {pages} {21} (\bibinfo {year} {2022})}\BibitemShut
  {NoStop}%
\bibitem [{\citenamefont {Bell}(1966)}]{bell1966problem}%
  \BibitemOpen
  \bibfield  {author} {\bibinfo {author} {\bibfnamefont {J.~S.}\ \bibnamefont
  {Bell}},\ }\href {\doibase https://doi.org/10.1103/RevModPhys.38.447}
  {\bibfield  {journal} {\bibinfo  {journal} {Reviews of Modern physics}\
  }\textbf {\bibinfo {volume} {38}},\ \bibinfo {pages} {447} (\bibinfo {year}
  {1966})}\BibitemShut {NoStop}%
\bibitem [{\citenamefont {Kochen}\ and\ \citenamefont
  {Specker}(1967)}]{kochen1967problem}%
  \BibitemOpen
  \bibfield  {author} {\bibinfo {author} {\bibfnamefont {S.}~\bibnamefont
  {Kochen}}\ and\ \bibinfo {author} {\bibfnamefont {E.~P.}\ \bibnamefont
  {Specker}},\ }\href {http://www.jstor.org/stable/24902153} {\bibfield
  {journal} {\bibinfo  {journal} {Journal of Mathematics and Mechanics}\
  }\textbf {\bibinfo {volume} {17}},\ \bibinfo {pages} {59} (\bibinfo {year}
  {1967})}\BibitemShut {NoStop}%
\bibitem [{\citenamefont {Cabello}\ and\ \citenamefont
  {Garc{\'\i}a-Alcaine}(1996)}]{cabello1996bell}%
  \BibitemOpen
  \bibfield  {author} {\bibinfo {author} {\bibfnamefont {A.}~\bibnamefont
  {Cabello}}\ and\ \bibinfo {author} {\bibfnamefont {G.}~\bibnamefont
  {Garc{\'\i}a-Alcaine}},\ }\href {\doibase 10.1088/0305-4470/29/5/016}
  {\bibfield  {journal} {\bibinfo  {journal} {Journal of Physics A:
  Mathematical and General}\ }\textbf {\bibinfo {volume} {29}},\ \bibinfo
  {pages} {1025} (\bibinfo {year} {1996})}\BibitemShut {NoStop}%
\bibitem [{\citenamefont {Peres}(1991)}]{peres1991two}%
  \BibitemOpen
  \bibfield  {author} {\bibinfo {author} {\bibfnamefont {A.}~\bibnamefont
  {Peres}},\ }\href {\doibase 10.1088/0305-4470/24/4/003} {\bibfield  {journal}
  {\bibinfo  {journal} {Journal of Physics A: Mathematical and General}\
  }\textbf {\bibinfo {volume} {24}},\ \bibinfo {pages} {L175} (\bibinfo {year}
  {1991})}\BibitemShut {NoStop}%
\bibitem [{\citenamefont {Peres}(2002)}]{Peres2002}%
  \BibitemOpen
  \bibinfo {editor} {\bibfnamefont {A.}~\bibnamefont {Peres}},\ ed.,\ \enquote
  {\bibinfo {title} {Contextuality},}\ in\ \href {\doibase
  10.1007/0-306-47120-5_7} {\emph {\bibinfo {booktitle} {Quantum Theory:
  Concepts and Methods}}}\ (\bibinfo  {publisher} {Springer Netherlands},\
  \bibinfo {address} {Dordrecht},\ \bibinfo {year} {2002})\ pp.\ \bibinfo
  {pages} {187--212}\BibitemShut {NoStop}%
\bibitem [{\citenamefont {Mermin}(1990)}]{mermin1990simple}%
  \BibitemOpen
  \bibfield  {author} {\bibinfo {author} {\bibfnamefont {N.~D.}\ \bibnamefont
  {Mermin}},\ }\href {\doibase https://doi.org/10.1103/PhysRevLett.65.3373}
  {\bibfield  {journal} {\bibinfo  {journal} {Physical review letters}\
  }\textbf {\bibinfo {volume} {65}},\ \bibinfo {pages} {3373} (\bibinfo {year}
  {1990})}\BibitemShut {NoStop}%
\bibitem [{\citenamefont {Mermin}(1993)}]{mermin1993hidden}%
  \BibitemOpen
  \bibfield  {author} {\bibinfo {author} {\bibfnamefont {N.~D.}\ \bibnamefont
  {Mermin}},\ }\href {\doibase https://doi.org/10.1103/RevModPhys.65.803}
  {\bibfield  {journal} {\bibinfo  {journal} {Reviews of Modern Physics}\
  }\textbf {\bibinfo {volume} {65}},\ \bibinfo {pages} {803} (\bibinfo {year}
  {1993})}\BibitemShut {NoStop}%
\bibitem [{\citenamefont {Abramsky}\ and\ \citenamefont
  {Brandenburger}(2011)}]{abramsky2011sheaf}%
  \BibitemOpen
  \bibfield  {author} {\bibinfo {author} {\bibfnamefont {S.}~\bibnamefont
  {Abramsky}}\ and\ \bibinfo {author} {\bibfnamefont {A.}~\bibnamefont
  {Brandenburger}},\ }\href {\doibase 10.1088/1367-2630/13/11/113036}
  {\bibfield  {journal} {\bibinfo  {journal} {New Journal of Physics}\ }\textbf
  {\bibinfo {volume} {13}},\ \bibinfo {pages} {113036} (\bibinfo {year}
  {2011})}\BibitemShut {NoStop}%
\bibitem [{\citenamefont {Bravyi}\ \emph {et~al.}(2017)\citenamefont {Bravyi},
  \citenamefont {Gambetta}, \citenamefont {Mezzacapo},\ and\ \citenamefont
  {Temme}}]{bravyi2017tapering}%
  \BibitemOpen
  \bibfield  {author} {\bibinfo {author} {\bibfnamefont {S.}~\bibnamefont
  {Bravyi}}, \bibinfo {author} {\bibfnamefont {J.~M.}\ \bibnamefont
  {Gambetta}}, \bibinfo {author} {\bibfnamefont {A.}~\bibnamefont {Mezzacapo}},
  \ and\ \bibinfo {author} {\bibfnamefont {K.}~\bibnamefont {Temme}},\
  }\href@noop {} {\bibfield  {journal} {\bibinfo  {journal} {arXiv preprint
  arXiv:1701.08213}\ } (\bibinfo {year} {2017})}\BibitemShut {NoStop}%
\bibitem [{\citenamefont {Izmaylov}\ \emph {et~al.}(2019)\citenamefont
  {Izmaylov}, \citenamefont {Yen}, \citenamefont {Lang},\ and\ \citenamefont
  {Verteletskyi}}]{izmaylov2019unitary}%
  \BibitemOpen
  \bibfield  {author} {\bibinfo {author} {\bibfnamefont {A.~F.}\ \bibnamefont
  {Izmaylov}}, \bibinfo {author} {\bibfnamefont {T.-C.}\ \bibnamefont {Yen}},
  \bibinfo {author} {\bibfnamefont {R.~A.}\ \bibnamefont {Lang}}, \ and\
  \bibinfo {author} {\bibfnamefont {V.}~\bibnamefont {Verteletskyi}},\
  }\href@noop {} {\bibfield  {journal} {\bibinfo  {journal} {Journal of
  chemical theory and computation}\ }\textbf {\bibinfo {volume} {16}},\
  \bibinfo {pages} {190} (\bibinfo {year} {2019})}\BibitemShut {NoStop}%
\bibitem [{\citenamefont {Zhao}\ \emph {et~al.}(2020)\citenamefont {Zhao},
  \citenamefont {Tranter}, \citenamefont {Kirby}, \citenamefont {Ung},
  \citenamefont {Miyake},\ and\ \citenamefont {Love}}]{zhao2020measurement}%
  \BibitemOpen
  \bibfield  {author} {\bibinfo {author} {\bibfnamefont {A.}~\bibnamefont
  {Zhao}}, \bibinfo {author} {\bibfnamefont {A.}~\bibnamefont {Tranter}},
  \bibinfo {author} {\bibfnamefont {W.~M.}\ \bibnamefont {Kirby}}, \bibinfo
  {author} {\bibfnamefont {S.~F.}\ \bibnamefont {Ung}}, \bibinfo {author}
  {\bibfnamefont {A.}~\bibnamefont {Miyake}}, \ and\ \bibinfo {author}
  {\bibfnamefont {P.~J.}\ \bibnamefont {Love}},\ }\href@noop {} {\bibfield
  {journal} {\bibinfo  {journal} {Physical {R}eview A}\ }\textbf {\bibinfo
  {volume} {101}},\ \bibinfo {pages} {062322} (\bibinfo {year}
  {2020})}\BibitemShut {NoStop}%
\bibitem [{\citenamefont {Raussendorf}\ \emph {et~al.}(2020)\citenamefont
  {Raussendorf}, \citenamefont {Bermejo-Vega}, \citenamefont {Tyhurst},
  \citenamefont {Okay},\ and\ \citenamefont {Zurel}}]{raussendorf2020phase}%
  \BibitemOpen
  \bibfield  {author} {\bibinfo {author} {\bibfnamefont {R.}~\bibnamefont
  {Raussendorf}}, \bibinfo {author} {\bibfnamefont {J.}~\bibnamefont
  {Bermejo-Vega}}, \bibinfo {author} {\bibfnamefont {E.}~\bibnamefont
  {Tyhurst}}, \bibinfo {author} {\bibfnamefont {C.}~\bibnamefont {Okay}}, \
  and\ \bibinfo {author} {\bibfnamefont {M.}~\bibnamefont {Zurel}},\
  }\href@noop {} {\bibfield  {journal} {\bibinfo  {journal} {Physical {R}eview
  A}\ }\textbf {\bibinfo {volume} {101}},\ \bibinfo {pages} {012350} (\bibinfo
  {year} {2020})}\BibitemShut {NoStop}%
\bibitem [{\citenamefont {Ralli}\ \emph {et~al.}(2021)\citenamefont {Ralli},
  \citenamefont {Love}, \citenamefont {Tranter},\ and\ \citenamefont
  {Coveney}}]{ralli2021implementation}%
  \BibitemOpen
  \bibfield  {author} {\bibinfo {author} {\bibfnamefont {A.}~\bibnamefont
  {Ralli}}, \bibinfo {author} {\bibfnamefont {P.~J.}\ \bibnamefont {Love}},
  \bibinfo {author} {\bibfnamefont {A.}~\bibnamefont {Tranter}}, \ and\
  \bibinfo {author} {\bibfnamefont {P.~V.}\ \bibnamefont {Coveney}},\
  }\href@noop {} {\bibfield  {journal} {\bibinfo  {journal} {Physical {R}eview
  Research}\ }\textbf {\bibinfo {volume} {3}},\ \bibinfo {pages} {033195}
  (\bibinfo {year} {2021})}\BibitemShut {NoStop}%
\bibitem [{\citenamefont {McCrimmon}(2004)}]{mccrimmon2004taste}%
  \BibitemOpen
  \bibfield  {author} {\bibinfo {author} {\bibfnamefont {K.}~\bibnamefont
  {McCrimmon}},\ }\href@noop {} {\emph {\bibinfo {title} {A taste of Jordan
  algebras}}},\ Vol.~\bibinfo {volume} {1}\ (\bibinfo  {publisher} {Springer},\
  \bibinfo {year} {2004})\BibitemShut {NoStop}%
\bibitem [{\citenamefont {Jansen}\ \emph {et~al.}(1997)\citenamefont {Jansen},
  \citenamefont {Scheffler},\ and\ \citenamefont
  {Woeginger}}]{jansen1997disjoint}%
  \BibitemOpen
  \bibfield  {author} {\bibinfo {author} {\bibfnamefont {K.}~\bibnamefont
  {Jansen}}, \bibinfo {author} {\bibfnamefont {P.}~\bibnamefont {Scheffler}}, \
  and\ \bibinfo {author} {\bibfnamefont {G.}~\bibnamefont {Woeginger}},\
  }\href@noop {} {\bibfield  {journal} {\bibinfo  {journal} {RAIRO-Operations
  Research}\ }\textbf {\bibinfo {volume} {31}},\ \bibinfo {pages} {45}
  (\bibinfo {year} {1997})}\BibitemShut {NoStop}%
\bibitem [{\citenamefont {Temme}\ and\ \citenamefont
  {Kastoryano}(2015)}]{temme2015fast}%
  \BibitemOpen
  \bibfield  {author} {\bibinfo {author} {\bibfnamefont {K.}~\bibnamefont
  {Temme}}\ and\ \bibinfo {author} {\bibfnamefont {M.~J.}\ \bibnamefont
  {Kastoryano}},\ }\href@noop {} {\bibfield  {journal} {\bibinfo  {journal}
  {arXiv preprint arXiv:1505.07811}\ } (\bibinfo {year} {2015})}\BibitemShut
  {NoStop}%
\bibitem [{\citenamefont {Kitaev}(1995)}]{kitaev1995quantum}%
  \BibitemOpen
  \bibfield  {author} {\bibinfo {author} {\bibfnamefont {A.~Y.}\ \bibnamefont
  {Kitaev}},\ }\href@noop {} {\bibfield  {journal} {\bibinfo  {journal} {arXiv
  preprint quant-ph/9511026}\ } (\bibinfo {year} {1995})}\BibitemShut {NoStop}%
\bibitem [{\citenamefont {Peruzzo}\ \emph {et~al.}(2014)\citenamefont
  {Peruzzo}, \citenamefont {McClean}, \citenamefont {Shadbolt}, \citenamefont
  {Yung}, \citenamefont {Zhou}, \citenamefont {Love}, \citenamefont
  {Aspuru-Guzik},\ and\ \citenamefont {O’brien}}]{peruzzo2014variational}%
  \BibitemOpen
  \bibfield  {author} {\bibinfo {author} {\bibfnamefont {A.}~\bibnamefont
  {Peruzzo}}, \bibinfo {author} {\bibfnamefont {J.}~\bibnamefont {McClean}},
  \bibinfo {author} {\bibfnamefont {P.}~\bibnamefont {Shadbolt}}, \bibinfo
  {author} {\bibfnamefont {M.-H.}\ \bibnamefont {Yung}}, \bibinfo {author}
  {\bibfnamefont {X.-Q.}\ \bibnamefont {Zhou}}, \bibinfo {author}
  {\bibfnamefont {P.~J.}\ \bibnamefont {Love}}, \bibinfo {author}
  {\bibfnamefont {A.}~\bibnamefont {Aspuru-Guzik}}, \ and\ \bibinfo {author}
  {\bibfnamefont {J.~L.}\ \bibnamefont {O’brien}},\ }\href {\doibase
  https://doi.org/10.1038/ncomms5213} {\bibfield  {journal} {\bibinfo
  {journal} {Nature communications}\ }\textbf {\bibinfo {volume} {5}},\
  \bibinfo {pages} {4213} (\bibinfo {year} {2014})}\BibitemShut {NoStop}%
\bibitem [{\citenamefont {Ravi}\ \emph {et~al.}(2022)\citenamefont {Ravi},
  \citenamefont {Gokhale}, \citenamefont {Ding}, \citenamefont {Kirby},
  \citenamefont {Smith}, \citenamefont {Baker}, \citenamefont {Love},
  \citenamefont {Hoffmann}, \citenamefont {Brown},\ and\ \citenamefont
  {Chong}}]{ravi2022cafqa}%
  \BibitemOpen
  \bibfield  {author} {\bibinfo {author} {\bibfnamefont {G.~S.}\ \bibnamefont
  {Ravi}}, \bibinfo {author} {\bibfnamefont {P.}~\bibnamefont {Gokhale}},
  \bibinfo {author} {\bibfnamefont {Y.}~\bibnamefont {Ding}}, \bibinfo {author}
  {\bibfnamefont {W.}~\bibnamefont {Kirby}}, \bibinfo {author} {\bibfnamefont
  {K.}~\bibnamefont {Smith}}, \bibinfo {author} {\bibfnamefont {J.~M.}\
  \bibnamefont {Baker}}, \bibinfo {author} {\bibfnamefont {P.~J.}\ \bibnamefont
  {Love}}, \bibinfo {author} {\bibfnamefont {H.}~\bibnamefont {Hoffmann}},
  \bibinfo {author} {\bibfnamefont {K.~R.}\ \bibnamefont {Brown}}, \ and\
  \bibinfo {author} {\bibfnamefont {F.~T.}\ \bibnamefont {Chong}},\ }in\ \href
  {https://doi.org/10.1145/3567955.3567958} {\emph {\bibinfo {booktitle}
  {Proceedings of the 28th ACM International Conference on Architectural
  Support for Programming Languages and Operating Systems, Volume 1}}}\
  (\bibinfo {year} {2022})\ pp.\ \bibinfo {pages} {15--29}\BibitemShut
  {NoStop}%
\bibitem [{\citenamefont {Aharonov}\ and\ \citenamefont
  {Ta-Shma}(2003)}]{aharonov2003adiabatic}%
  \BibitemOpen
  \bibfield  {author} {\bibinfo {author} {\bibfnamefont {D.}~\bibnamefont
  {Aharonov}}\ and\ \bibinfo {author} {\bibfnamefont {A.}~\bibnamefont
  {Ta-Shma}},\ }in\ \href@noop {} {\emph {\bibinfo {booktitle} {Proceedings of
  the thirty-fifth annual ACM symposium on Theory of computing}}}\ (\bibinfo
  {year} {2003})\ pp.\ \bibinfo {pages} {20--29},\ \Eprint
  {http://arxiv.org/abs/quant-ph/0301023} {arXiv:quant-ph/0301023} \BibitemShut
  {NoStop}%
\bibitem [{\citenamefont {Albash}\ and\ \citenamefont
  {Lidar}(2018)}]{albash2018adiabatic}%
  \BibitemOpen
  \bibfield  {author} {\bibinfo {author} {\bibfnamefont {T.}~\bibnamefont
  {Albash}}\ and\ \bibinfo {author} {\bibfnamefont {D.~A.}\ \bibnamefont
  {Lidar}},\ }\href@noop {} {\bibfield  {journal} {\bibinfo  {journal}
  {{R}eviews of Modern Physics}\ }\textbf {\bibinfo {volume} {90}},\ \bibinfo
  {pages} {015002} (\bibinfo {year} {2018})}\BibitemShut {NoStop}%
\bibitem [{\citenamefont {Patel}\ \emph {et~al.}(2023)\citenamefont {Patel},
  \citenamefont {Yen},\ and\ \citenamefont {Izmaylov}}]{patel2023extension}%
  \BibitemOpen
  \bibfield  {author} {\bibinfo {author} {\bibfnamefont {S.}~\bibnamefont
  {Patel}}, \bibinfo {author} {\bibfnamefont {T.-C.}\ \bibnamefont {Yen}}, \
  and\ \bibinfo {author} {\bibfnamefont {A.~F.}\ \bibnamefont {Izmaylov}},\
  }\href@noop {} {\bibfield  {journal} {\bibinfo  {journal} {arXiv}\ }
  (\bibinfo {year} {2023})},\ \Eprint {http://arxiv.org/abs/2305.18251}
  {2305.18251} \BibitemShut {NoStop}%
\bibitem [{\citenamefont {Patel}\ and\ \citenamefont
  {Izmaylov}(2024)}]{patel2024exactly}%
  \BibitemOpen
  \bibfield  {author} {\bibinfo {author} {\bibfnamefont {S.}~\bibnamefont
  {Patel}}\ and\ \bibinfo {author} {\bibfnamefont {A.~F.}\ \bibnamefont
  {Izmaylov}},\ }\href@noop {} {\bibfield  {journal} {\bibinfo  {journal} {The
  Journal of Chemical Physics}\ }\textbf {\bibinfo {volume} {160}} (\bibinfo
  {year} {2024})}\BibitemShut {NoStop}%
\end{thebibliography}%

\onecolumngrid
\appendix
\addcontentsline{toc}{section}{Appendices}
\clearpage
\section*{Appendix}
\section{Size of Noncontextual Pauli Hamiltonian} \label{sec:nonconHSize}
Table \ref{tab:fine_grain_nonconSize} gives the number of Pauli operators a noncontextual Pauli Hamiltonian according to the size of $\mathcal{G}$ and $\mathcal{A}$, as defined by the first two lines of equation \ref{eq:NonconTerms} (the other lines require $|\mathcal{A}|$ to have maximum size). Note all noncontextual Hamiltonians with $|\mathcal{A}|>8$ must have fewer than $2^{n}$ Pauli operators, which can be determined from equation \ref{eq:NonconTerms}. As commented on in the main text, this analysis is given to show how expressive $H_{nc}$ can be. In practice, only Hamiltonians described on a polynomial number of Pauli operators can be studied.

\begin{table}[h]
\begin{tabular}{c|c|cc}
\hline
\hline
$|G|$ & $| A |$ & max $|H_{nc}|$                                                 & max $|H_{nc}| > 2^{n}$ \\ \hline \hline
$n$   & 0       & $2^{n}$                                                        & EQUAL         \\ \hline
$n-1$ & 0       & $2^{n-1}$                                                      & N             \\
$n-1$ & 2       & $2^{n}+2^{n-1}$            & Y             \\
$n-1$ & 3       & $2^{n+1}$                                                      & Y             \\ \hline
$n-2$ & 0       & $2^{n-2}$                                                      & N             \\
$n-2$ & 2       & $2^{n-1}+2^{n-2}$          & N             \\
$n-2$ & 3       & $ 2^{n}$                                                       & EQUAL         \\
$n-2$ & 4       & $ 2^{n}+2^{n-2}$           & Y             \\
$n-2$ & 5       & $2^{n}+2^{n-1}$         & Y             \\ \hline
$n-3$ & 0       & $2^{n-3}$                                                      & N             \\
$n-3$ & 2       & $ 2^{n-2}+2^{n-3}$          & N             \\
$n-3$ & 3       & $ 2^{n-1}$                                  & N             \\
$n-3$ & 4       & $ 2^{n-1}+2^{n-3}$          & N         \\
$n-3$ & 5       & $ 2^{n-1}+2^{n-2}$         & N             \\
$n-3$ & 6       & $ 2^{n-1}+2^{n-2}+2^{n-3}$  & N             \\
$n-3$ & 7       & $ 2^{n}$                                    & EQUAL             \\ \hline
\end{tabular}
\caption{Maximum number of Pauli operators a noncontextual Pauli Hamiltonian $H_{\text{nc}}$ can be composed of compared to size of $\mathcal{G}$ and $\mathcal{A}$. In the final column, $Y$ indicates yes the maximum support of $H_{nc}$ can be greater than $2^{n}$  and $N$ indicates no the support of $H_{nc}$ must be less than $2^{n}$. }
    \label{tab:fine_grain_nonconSize}
\end{table}

\section{Expectation value of a linear combination of anticommuting Pauli operators} \label{sec:AC_exp}

In this Appendix we prove Theorem \ref{AntiCommExpec} in the main text. 

\textit{Proof.} First, consider a general real linear combination of pairwise anticommuting Pauli operators acting on $n$-qubits:

\begin{equation}
\label{eq:AntiSum2}
     D = \sum_{\substack{k=1 \\ P_{k} \in \mathcal{V}_{D }}}^{|\mathcal{V}_{D}|\leq 2n+1} c_{k} P_{k},
\end{equation}
where:

\begin{subequations}
\begin{equation}
\label{eq:antideff}
  \{P_{a}, P_{b}\}=0 \: \: \forall a \neq b \text{ and } P_{a}, P_{b} \in \mathcal{V}_{D},
\end{equation}    
\begin{equation}
  c_{k} \in \mathbb{R}.
\end{equation}
\end{subequations}

Let us normalize the sum in equation \ref{eq:AntiSum2} by the euclidean norm of the coefficients in the linear combination:

\begin{equation}
\label{eq:AntiSum3}
     D = \big(\sum_{k}c_{k}^{2}\big)^{1/2}\sum_{\substack{k=1 \\ P_{k} \in \mathcal{V}_{D }}}^{|\mathcal{V}_{D}|\leq 2n+1} \beta_{k} P_{k} = {\|\vec{c}\|_{2}}\sum_{\substack{k=1 \\ P_{k} \in \mathcal{V}_{D }}}^{|\mathcal{V}_{D}|\leq 2n+1} \beta_{k} P_{k},
\end{equation}
where:
\begin{equation}
     \beta_{j} = \frac{c_{j}}{(\sum_{k}c_{k}^{2})^{1/2}}= \frac{c_{j}}{\|\vec{c}\|_{2}}.
\end{equation}
Theorem \ref{UP_map} and equation \ref{eq:UP_summarry} allow us to  transform equation \ref{eq:AntiSum3} as:

\begin{equation}
\label{eq:AntiSum4}
\begin{aligned}
      D &= \|\vec{c}\|_{2} \sum_{\substack{k=1 \\ P_{k} \in \mathcal{V}_{D }}}^{|\mathcal{V}_{D}|\leq 2n+1} \beta_{k} P_{k} \\
     &\mapsto \\
      R^{\dagger}D R  &= \|\vec{c}\|_{2} \; R^{\dagger} \bigg( \sum_{\substack{k=1 \\ P_{k} \in \mathcal{V}_{D }}}^{|\mathcal{V}_{D}|\leq 2n+1} \beta_{k} P_{k}\bigg) R  = \|\vec{c}\|_{2} \; P.
     \end{aligned}
\end{equation}
Here $R$ is a unitary operator defined as either $R_{\text{LCU}}$ or $R_{\text{S}}$ (see Subsection \ref{sec:UP}), therfore $\|\vec{c}\|_{2} P$ must be isospectral to $D$. By Corollary \ref{single_Q_map}, a Clifford (unitary) operator can be constructed to map this $n$-qubit Pauli operator $P$ to a single Pauli $Z_{j}$ acting on a unique qubit with index $j$. We write this as: 

\begin{equation}
\label{eq:AntiSum5}
\|\vec{c}\|_{2} \;P  \mapsto  \|\vec{c}\|_{2} \big( I_{1} \otimes \hdots \otimes I_{j-1} \otimes Z_{j} \otimes I_{j+1}\otimes \hdots  \otimes I_{n} \big).
\end{equation}
The resulting Pauli operator, which acts non-trivially only on qubit $j$, has eigenvalue $+\|\vec{c}\|_{2}$ with eigenvectors $I_{1} \otimes \hdots \otimes I_{j-1} \otimes \ket{0}_{j} \otimes I_{j+1} \otimes \hdots \otimes I_{n}$ and eigenvalue $-\|\vec{c}\|_{2}$ with eigenvectors $I_{1} \otimes \hdots \otimes I_{j-1} \otimes \ket{1}_{j} \otimes I_{j+1}\otimes \hdots  \otimes I_{n}$. Clearly this operator has a degenerate spectrum and the number of times each eigenvalue occurs will be $2^{n-1}$, due to the $n-1$ identity terms. As all transformations in this proof are unitary, $D$ must have the same eigenvalues as $\|\vec{c}\|_{2} \big( I_{1} \otimes \hdots \otimes I_{j-1} \otimes Z_{j} \otimes I_{j+1} \otimes \hdots \otimes I_{n} \big)$ \qedsymbol{}.

\section{Classical Evaluation of the Noncontextual Eigenvalues}\label{noncon_obj_func}

The noncontextual Hamiltonian can be written
\begin{equation}
    H_{\mathrm{nc}} = S_0 + \sum_{i=1}^M C_i S_i
\end{equation}
where $S_i \coloneqq \sum_{P \in \overline{\mathcal{G}}} h_P^{(i)} P^{(i)}$ are Hamiltonian symmetries, whose Pauli terms are products of the symmetry generators $\mathcal{G}$, and the clique representatives $C_i$ are pairwise anticommuting, i.e. $\{C_i, C_j\} = 0 \;\forall i \neq j$. Such Hamiltonians lend themselves to a convenient description via a classical objective function, which we derive now. 

Firstly, since $\mathcal{G}$ forms a set of commuting Pauli operators we may simultaneously assign eigenvalues $G \mapsto \nu_G = \pm1$ for each $G_{i} \in\mathcal{G}$. In turn, this specifies expectation values of the symmetry contributions $S_i$, namely
\begin{equation}
    s_i(\vec{\nu}) \coloneqq \sum_{P \in \overline{\mathcal{G}}} h_P^{(i)} \prod_{G_{i} \in \mathcal{G}_P} \nu_G
\end{equation}
where $\mathcal{G}_P \subset \mathcal{G}$ is the unique subset of symmetry generators required to reconstruct the Pauli term $P$, i.e. $P = \prod_{G_{i} \in \mathcal{G}_P} G$. Having assigned a collection of eigenvalues $\vec{\nu} \in \{\pm1\}^{\times |\mathcal{G}|}$, we obtain
\begin{equation}\label{G_assigned}
    H_{\mathrm{nc}}(\vec{\nu}) = s_0(\vec{\nu}) I + \sum_{i=1}^M s_i(\vec{\nu}) C_i,
\end{equation}
leaving an identity term and a sum over anticommuting Pauli operators. Defining $\vec{s}(\vec{\nu}) \coloneqq \big(s_i(\vec{\nu})\big)_{i=1}^M$ the vector of symmetry contributions associated with some clique representative, we may rewrite \eqref{G_assigned} as
\begin{equation}\label{C_normalized}
    H_{\mathrm{nc}}(\vec{\nu}) = s_0(\vec{\nu}) I + ||\vec{s}(\vec{\nu})|| C(\vec{\nu})
\end{equation}
where $C(\vec{\nu}) \coloneqq \sum_{i=1}^M \frac{s_i(\vec{\nu})}{||\vec{s}(\vec{\nu})||} C_i$ is unitary since it consists of a sum over anticommuting operators and its coefficients are normalized. Furthermore, $C(\vec{\nu})$ is unitarily equivalent to a Pauli operator by the theory of \textit{unitary partitioning}, therefore it has eigenvalues $\pm1$ and we obtain the noncontextual energy
\begin{equation}\label{C_assigned}
\begin{aligned}
    H_{\mathrm{nc}}^{\pm}(\vec{\nu}) 
    ={} & s_0(\vec{\nu}) \pm ||\vec{s}(\vec{\nu})|| \\
    ={} & s_0(\vec{\nu}) \pm \sqrt{s_1^2(\vec{\nu}) + \dots + s_M^2(\vec{\nu})}.
\end{aligned}
\end{equation}
To identify the noncontextual ground state energy we optimize over eigenvalue assignments $\vec{\nu}$ (note we may always take the $-1$ eigenvalue for $C(\vec{\nu})$ since $||\vec{s}(\vec{\nu})|| \geq 0$ it will always yield a lower value):
\begin{equation}
\label{eq:noncon_min}
    \epsilon_0 = \min_{\vec{\nu} \in \{\pm1\}^{\times |\mathcal{G}|}} H_{\mathrm{nc}}^{-}(\vec{\nu}).
\end{equation}

\subsubsection{Extracting an Eigenstate for a Particular Noncontextual Eigenvalue}

The operators $\mathcal{G} \cup \{C(\vec{\nu})\}$ are commuting, hence they share a common eigenbasis. Furthermore, it is possible to define a unitary operation $U$ mapping $C(\vec{\nu})$ onto one of its constituent clique representatives, e.g. $U C(\vec{\nu}) U^\dag = C_1$ for example; this may be achieved efficiently via the theory of unitary partitioning. Therefore, to identify an eigenstate for some eigenvalue $H_{\mathrm{nc}}^{\pm}(\vec{\nu})$ of the noncontextual Hamiltonian, we select a state $\psi^\prime$ that is stabilized by the operators $\pm C_1$ and $\nu_G G$ for every $G_{i} \in \mathcal{G}$. Then, since $[U, G] = 0 \; \forall G_{i} \in \mathcal{G}$, we have that $\psi \coloneqq U \psi^\prime$ is stabilized by the operators $\{\nu_G G\}_{G_{i} \in \mathcal{G}} \cup \{\pm C(\vec{\nu})\}$. Finally, using equations \eqref{G_assigned} -- \eqref{C_assigned} we have
\begin{equation}
\begin{aligned}
    H_{\mathrm{nc}} \psi 
    ={} & \big(s_0(\vec{\nu}) \pm ||\vec{s}(\vec{\nu})||\big) \psi \\
    ={} & H_{\mathrm{nc}}^{\pm}(\vec{\nu}) \psi.
\end{aligned}
\end{equation}
This demonstrates that our noncontextual objective function \eqref{C_assigned} produces legitimate eigenvalues of $H_{\mathrm{nc}}$. It is now possible to use $\psi$ for further computation, for example as a reference state in the contextual subspace framework.
\end{document}